\documentclass[iop]{emulateapj}

\shorttitle{Near-infrared thermal emission from WASP-12b} 
\shortauthors{Croll et al.}

\newcommand{\FpOverFStarPercentAbstractWASPTwelveKs}{0.310}
\newcommand{\FpOverFStarPercentAbstractMinusWASPTwelveKs}{0.013}
\newcommand{\FpOverFStarPercentAbstractPlusWASPTwelveKs}{0.012}
\newcommand{\XSigmaWASPTwelveKs}{24}

\newcommand{\PhaseAbstractWASPTwelveKs}{0.4994}
\newcommand{\PhaseAbstractMinusWASPTwelveKs}{0.0008}
\newcommand{\PhaseAbstractPlusWASPTwelveKs}{0.0009}
\newcommand{\ECosOmegaWASPTwelveKs}{-0.0013}
\newcommand{\ECosOmegaPlusWASPTwelveKs}{0.0015}
\newcommand{\ECosOmegaMinusWASPTwelveKs}{0.0015}

\newcommand{\TBrightWASPTwelveKs}{2988}
\newcommand{\TBrightPlusWASPTwelveKs}{45}
\newcommand{\TBrightMinusWASPTwelveKs}{46}

\newcommand{\fReradiationWASPTwelveKs}{0.482}
\newcommand{\fReradiationPlusWASPTwelveKs}{0.030}
\newcommand{\fReradiationMinusWASPTwelveKs}{0.029}

\newcommand{\cOneWASPTwelveKs}{0.00116}
\newcommand{\cOnePlusWASPTwelveKs}{0.00014}
\newcommand{\cOneMinusWASPTwelveKs}{0.00014}
\newcommand{\cTwoWASPTwelveKs}{0.001}
\newcommand{\cTwoPlusWASPTwelveKs}{0.001}
\newcommand{\cTwoMinusWASPTwelveKs}{0.001}

\newcommand{\TOffsetWASPTwelveKs}{-1.3}

\newcommand{\TOffsetPlusWASPTwelveKs}{1.5}
\newcommand{\TOffsetMinusWASPTwelveKs}{1.2}

\newcommand{\EclipseDurationWASPTwelveKs}{2.93}

\newcommand{\ChiWASPTwelveKs}{0.732}
\newcommand{\ChiPlusWASPTwelveKs}{0.003}
\newcommand{\ChiMinusWASPTwelveKs}{0.001}
\newcommand{\JDOffsetONEWASPTwelveKs}{5194.9351}
\newcommand{\JDOffsetPlusONEWASPTwelveKs}{0.0010}
\newcommand{\JDOffsetMinusONEWASPTwelveKs}{0.0008}

\newcommand{\FpOverFStarPercentAbstractWASPTwelveH}{0.180}
\newcommand{\FpOverFStarPercentAbstractMinusWASPTwelveH}{0.018}
\newcommand{\FpOverFStarPercentAbstractPlusWASPTwelveH}{0.015}

\newcommand{\PhaseAbstractWASPTwelveH}{0.5015}
\newcommand{\PhaseAbstractMinusWASPTwelveH}{0.0017}
\newcommand{\PhaseAbstractPlusWASPTwelveH}{0.0013}
\newcommand{\ECosOmegaWASPTwelveH}{0.0020}
\newcommand{\ECosOmegaPlusWASPTwelveH}{0.0021}
\newcommand{\ECosOmegaMinusWASPTwelveH}{0.0021}

\newcommand{\TBrightWASPTwelveH}{2765}
\newcommand{\TBrightPlusWASPTwelveH}{70}
\newcommand{\TBrightMinusWASPTwelveH}{72}

\newcommand{\fReradiationWASPTwelveH}{0.353}
\newcommand{\fReradiationPlusWASPTwelveH}{0.037}
\newcommand{\fReradiationMinusWASPTwelveH}{0.036}

\newcommand{\cOneWASPTwelveH}{0.00109}
\newcommand{\cOnePlusWASPTwelveH}{0.00022}
\newcommand{\cOneMinusWASPTwelveH}{0.00021}
\newcommand{\cTwoWASPTwelveH}{-0.003}
\newcommand{\cTwoPlusWASPTwelveH}{0.001}
\newcommand{\cTwoMinusWASPTwelveH}{0.001}

\newcommand{\TOffsetWASPTwelveH}{2.0}

\newcommand{\TOffsetPlusWASPTwelveH}{2.1}
\newcommand{\TOffsetMinusWASPTwelveH}{2.6}

\newcommand{\EclipseDurationWASPTwelveH}{2.93}

\newcommand{\ChiWASPTwelveH}{0.435}
\newcommand{\ChiPlusWASPTwelveH}{0.002}
\newcommand{\ChiMinusWASPTwelveH}{0.001}
\newcommand{\JDOffsetONEWASPTwelveH}{5193.8461}
\newcommand{\JDOffsetPlusONEWASPTwelveH}{0.0014}
\newcommand{\JDOffsetMinusONEWASPTwelveH}{0.0018}

\newcommand{\FpOverFStarPercentAbstractWASPTwelveJ}{0.126}

\newcommand{\FpOverFStarPercentAbstractPlusWASPTwelveJ}{0.030}

\newcommand{\PhaseAbstractWASPTwelveJ}{0.4987}
\newcommand{\PhaseAbstractMinusWASPTwelveJ}{0.0028}
\newcommand{\PhaseAbstractPlusWASPTwelveJ}{0.0028}
\newcommand{\ECosOmegaWASPTwelveJ}{-0.0024}
\newcommand{\ECosOmegaPlusWASPTwelveJ}{0.0044}
\newcommand{\ECosOmegaMinusWASPTwelveJ}{0.0044}

\newcommand{\TBrightWASPTwelveJ}{2833}
\newcommand{\TBrightPlusWASPTwelveJ}{152}
\newcommand{\TBrightMinusWASPTwelveJ}{173}

\newcommand{\fReradiationWASPTwelveJ}{0.389}
\newcommand{\fReradiationPlusWASPTwelveJ}{0.091}
\newcommand{\fReradiationMinusWASPTwelveJ}{0.087}

\newcommand{\cOneWASPTwelveJ}{0.00157}
\newcommand{\cOnePlusWASPTwelveJ}{0.00027}
\newcommand{\cOneMinusWASPTwelveJ}{0.00035}
\newcommand{\cTwoWASPTwelveJ}{-0.014}
\newcommand{\cTwoPlusWASPTwelveJ}{0.003}
\newcommand{\cTwoMinusWASPTwelveJ}{0.003}

\newcommand{\TOffsetWASPTwelveJ}{-2.4}

\newcommand{\TOffsetPlusWASPTwelveJ}{4.5}
\newcommand{\TOffsetMinusWASPTwelveJ}{4.5}

\newcommand{\EclipseDurationWASPTwelveJ}{2.93}

\newcommand{\ChiWASPTwelveJ}{0.358}
\newcommand{\ChiPlusWASPTwelveJ}{0.003}
\newcommand{\ChiMinusWASPTwelveJ}{0.002}
\newcommand{\JDOffsetONEWASPTwelveJ}{5192.7515}
\newcommand{\JDOffsetPlusONEWASPTwelveJ}{0.0031}
\newcommand{\JDOffsetMinusONEWASPTwelveJ}{0.0031}

\newcommand{\FpOverFStarPercentAbstractWASPTwelveJointAll}{0.309}
\newcommand{\FpOverFStarPercentAbstractMinusWASPTwelveJointAll}{0.012}
\newcommand{\FpOverFStarPercentAbstractPlusWASPTwelveJointAll}{0.013}
\newcommand{\XSigmaWASPTwelveJointAll}{24}

\newcommand{\PhaseAbstractWASPTwelveJointAll}{0.4998}
\newcommand{\PhaseAbstractMinusWASPTwelveJointAll}{0.0007}
\newcommand{\PhaseAbstractPlusWASPTwelveJointAll}{0.0008}
\newcommand{\ECosOmegaWASPTwelveJointAll}{-0.0007}
\newcommand{\ECosOmegaPlusWASPTwelveJointAll}{0.0013}
\newcommand{\ECosOmegaMinusWASPTwelveJointAll}{0.0013}
\newcommand{\ECosOmegaAbsoluteThreeSigmaLimitWASPTwelveJointAll}{0.0040}
\newcommand{\TBrightWASPTwelveJointAll}{2985}
\newcommand{\TBrightPlusWASPTwelveJointAll}{49}
\newcommand{\TBrightMinusWASPTwelveJointAll}{49}

\newcommand{\fReradiationWASPTwelveJointAll}{0.480}
\newcommand{\fReradiationPlusWASPTwelveJointAll}{0.032}
\newcommand{\fReradiationMinusWASPTwelveJointAll}{0.031}

\newcommand{\cOneWASPTwelveJointAll}{0.00116}
\newcommand{\cOnePlusWASPTwelveJointAll}{0.00012}
\newcommand{\cOneMinusWASPTwelveJointAll}{0.00015}
\newcommand{\cTwoWASPTwelveJointAll}{0.001}
\newcommand{\cTwoPlusWASPTwelveJointAll}{0.001}
\newcommand{\cTwoMinusWASPTwelveJointAll}{0.001}

\newcommand{\TOffsetWASPTwelveJointAll}{-0.7}

\newcommand{\TOffsetPlusWASPTwelveJointAll}{1.3}
\newcommand{\TOffsetMinusWASPTwelveJointAll}{1.1}

\newcommand{\EclipseDurationWASPTwelveJointAll}{2.93}

\newcommand{\ChiWASPTwelveJointAll}{0.533}
\newcommand{\ChiPlusWASPTwelveJointAll}{0.002}
\newcommand{\ChiMinusWASPTwelveJointAll}{0.001}
\newcommand{\JDOffsetONEWASPTwelveJointAll}{5192.7527}
\newcommand{\JDOffsetPlusONEWASPTwelveJointAll}{0.0009}
\newcommand{\JDOffsetMinusONEWASPTwelveJointAll}{0.0007}
\newcommand{\JDOffsetTWOWASPTwelveJointAll}{5193.8441}
\newcommand{\JDOffsetPlusTWOWASPTwelveJointAll}{0.0009}
\newcommand{\JDOffsetMinusTWOWASPTwelveJointAll}{0.0007}
\newcommand{\JDOffsetTHREEWASPTwelveJointAll}{5194.9356}
\newcommand{\JDOffsetPlusTHREEWASPTwelveJointAll}{0.0009}
\newcommand{\JDOffsetMinusTHREEWASPTwelveJointAll}{0.0007}
\newcommand{\XSigmaSixWASPTwelveJointAll}{9}
\newcommand{\XSigmaNineWASPTwelveJointAll}{4}
\newcommand{\ParamSixWASPTwelveJointAll}{0.176}
\newcommand{\ParamSixPlusWASPTwelveJointAll}{0.016}
\newcommand{\ParamSixMinusWASPTwelveJointAll}{0.021}
\newcommand{\ParamSevenWASPTwelveJointAll}{0.00104}
\newcommand{\ParamSevenPlusWASPTwelveJointAll}{0.00024}
\newcommand{\ParamSevenMinusWASPTwelveJointAll}{0.00024}
\newcommand{\ParamEightWASPTwelveJointAll}{-0.003}
\newcommand{\ParamEightPlusWASPTwelveJointAll}{0.001}
\newcommand{\ParamEightMinusWASPTwelveJointAll}{0.001}
\newcommand{\ParamNineWASPTwelveJointAll}{0.131}
\newcommand{\ParamNinePlusWASPTwelveJointAll}{0.027}
\newcommand{\ParamNineMinusWASPTwelveJointAll}{0.029}
\newcommand{\ParamTenWASPTwelveJointAll}{0.00163}
\newcommand{\ParamTenPlusWASPTwelveJointAll}{0.00030}
\newcommand{\ParamTenMinusWASPTwelveJointAll}{0.00031}
\newcommand{\ParamElevenWASPTwelveJointAll}{-0.015}
\newcommand{\ParamElevenPlusWASPTwelveJointAll}{0.003}
\newcommand{\ParamElevenMinusWASPTwelveJointAll}{0.003}
\newcommand{\TBrightSixWASPTwelveJointAll}{2748}
\newcommand{\TBrightSixPlusWASPTwelveJointAll}{71}
\newcommand{\TBrightSixMinusWASPTwelveJointAll}{74}
\newcommand{\fReradiationSixWASPTwelveJointAll}{0.345}
\newcommand{\fReradiationSixPlusWASPTwelveJointAll}{0.037}
\newcommand{\fReradiationSixMinusWASPTwelveJointAll}{0.036}
\newcommand{\TBrightNineWASPTwelveJointAll}{2860}
\newcommand{\TBrightNinePlusWASPTwelveJointAll}{138}
\newcommand{\TBrightNineMinusWASPTwelveJointAll}{155}
\newcommand{\fReradiationNineWASPTwelveJointAll}{0.405}
\newcommand{\fReradiationNinePlusWASPTwelveJointAll}{0.085}
\newcommand{\fReradiationNineMinusWASPTwelveJointAll}{0.081}

\newcommand{\FpOverFStarPercentAbstractWASPTwelveVariableKs}{0.311}
\newcommand{\FpOverFStarPercentAbstractMinusWASPTwelveVariableKs}{0.011}
\newcommand{\FpOverFStarPercentAbstractPlusWASPTwelveVariableKs}{0.013}

\newcommand{\PhaseAbstractWASPTwelveVariableKs}{0.4997}
\newcommand{\PhaseAbstractMinusWASPTwelveVariableKs}{0.0009}
\newcommand{\PhaseAbstractPlusWASPTwelveVariableKs}{0.0009}
\newcommand{\ECosOmegaWASPTwelveVariableKs}{-0.0009}
\newcommand{\ECosOmegaPlusWASPTwelveVariableKs}{0.0014}
\newcommand{\ECosOmegaMinusWASPTwelveVariableKs}{0.0014}

\newcommand{\TBrightWASPTwelveVariableKs}{2993}
\newcommand{\TBrightPlusWASPTwelveVariableKs}{50}
\newcommand{\TBrightMinusWASPTwelveVariableKs}{51}

\newcommand{\fReradiationWASPTwelveVariableKs}{0.485}
\newcommand{\fReradiationPlusWASPTwelveVariableKs}{0.034}
\newcommand{\fReradiationMinusWASPTwelveVariableKs}{0.033}

\newcommand{\cOneWASPTwelveVariableKs}{0.00126}
\newcommand{\cOnePlusWASPTwelveVariableKs}{0.00016}
\newcommand{\cOneMinusWASPTwelveVariableKs}{0.00014}
\newcommand{\cTwoWASPTwelveVariableKs}{0.001}
\newcommand{\cTwoPlusWASPTwelveVariableKs}{0.001}
\newcommand{\cTwoMinusWASPTwelveVariableKs}{0.001}

\newcommand{\TOffsetWASPTwelveVariableKs}{-0.9}

\newcommand{\TOffsetPlusWASPTwelveVariableKs}{1.4}
\newcommand{\TOffsetMinusWASPTwelveVariableKs}{1.4}
\newcommand{\WidthFactorWASPTwelveVariableKs}{1.109}
\newcommand{\WidthFactorPlusWASPTwelveVariableKs}{0.046}
\newcommand{\WidthFactorMinusWASPTwelveVariableKs}{0.039}
\newcommand{\WidthFactorSigmaWASPTwelveVariableKs}{2.8}
\newcommand{\EclipseDurationWASPTwelveVariableKs}{3.25}
\newcommand{\EclipseDurationPlusWASPTwelveVariableKs}{0.14}
\newcommand{\EclipseDurationMinusWASPTwelveVariableKs}{0.11}
\newcommand{\ESinOmegaWASPTwelveVariableKs}{0.059}
\newcommand{\ESinOmegaPlusWASPTwelveVariableKs}{0.082}
\newcommand{\ESinOmegaMinusWASPTwelveVariableKs}{0.034}
\newcommand{\ChiWASPTwelveVariableKs}{0.727}
\newcommand{\ChiPlusWASPTwelveVariableKs}{0.003}
\newcommand{\ChiMinusWASPTwelveVariableKs}{0.001}
\newcommand{\JDOffsetONEWASPTwelveVariableKs}{5194.9355}
\newcommand{\JDOffsetPlusONEWASPTwelveVariableKs}{0.0010}
\newcommand{\JDOffsetMinusONEWASPTwelveVariableKs}{0.0010}

\newcommand{\FpOverFStarPercentAbstractWASPTwelveJointVariableAll}{0.313}
\newcommand{\FpOverFStarPercentAbstractMinusWASPTwelveJointVariableAll}{0.013}
\newcommand{\FpOverFStarPercentAbstractPlusWASPTwelveJointVariableAll}{0.011}

\newcommand{\PhaseAbstractWASPTwelveJointVariableAll}{0.4999}
\newcommand{\PhaseAbstractMinusWASPTwelveJointVariableAll}{0.0009}
\newcommand{\PhaseAbstractPlusWASPTwelveJointVariableAll}{0.0006}
\newcommand{\ECosOmegaWASPTwelveJointVariableAll}{-0.0006}
\newcommand{\ECosOmegaPlusWASPTwelveJointVariableAll}{0.0009}
\newcommand{\ECosOmegaMinusWASPTwelveJointVariableAll}{0.0009}

\newcommand{\TBrightWASPTwelveJointVariableAll}{3000}
\newcommand{\TBrightPlusWASPTwelveJointVariableAll}{40}
\newcommand{\TBrightMinusWASPTwelveJointVariableAll}{40}

\newcommand{\fReradiationWASPTwelveJointVariableAll}{0.490}
\newcommand{\fReradiationPlusWASPTwelveJointVariableAll}{0.027}
\newcommand{\fReradiationMinusWASPTwelveJointVariableAll}{0.026}

\newcommand{\cOneWASPTwelveJointVariableAll}{0.00122}
\newcommand{\cOnePlusWASPTwelveJointVariableAll}{0.00017}
\newcommand{\cOneMinusWASPTwelveJointVariableAll}{0.00013}
\newcommand{\cTwoWASPTwelveJointVariableAll}{0.001}
\newcommand{\cTwoPlusWASPTwelveJointVariableAll}{0.001}
\newcommand{\cTwoMinusWASPTwelveJointVariableAll}{0.001}

\newcommand{\TOffsetWASPTwelveJointVariableAll}{-0.6}

\newcommand{\TOffsetPlusWASPTwelveJointVariableAll}{0.9}
\newcommand{\TOffsetMinusWASPTwelveJointVariableAll}{1.4}
\newcommand{\WidthFactorWASPTwelveJointVariableAll}{1.080}
\newcommand{\WidthFactorPlusWASPTwelveJointVariableAll}{0.034}
\newcommand{\WidthFactorMinusWASPTwelveJointVariableAll}{0.034}

\newcommand{\EclipseDurationWASPTwelveJointVariableAll}{3.16}
\newcommand{\EclipseDurationPlusWASPTwelveJointVariableAll}{0.10}
\newcommand{\EclipseDurationMinusWASPTwelveJointVariableAll}{0.10}
\newcommand{\ESinOmegaWASPTwelveJointVariableAll}{0.044}
\newcommand{\ESinOmegaPlusWASPTwelveJointVariableAll}{0.062}
\newcommand{\ESinOmegaMinusWASPTwelveJointVariableAll}{0.026}
\newcommand{\ChiWASPTwelveJointVariableAll}{0.532}
\newcommand{\ChiPlusWASPTwelveJointVariableAll}{0.001}
\newcommand{\ChiMinusWASPTwelveJointVariableAll}{0.001}
\newcommand{\JDOffsetONEWASPTwelveJointVariableAll}{5192.7528}
\newcommand{\JDOffsetPlusONEWASPTwelveJointVariableAll}{0.0007}
\newcommand{\JDOffsetMinusONEWASPTwelveJointVariableAll}{0.0010}
\newcommand{\JDOffsetTWOWASPTwelveJointVariableAll}{5193.8442}
\newcommand{\JDOffsetPlusTWOWASPTwelveJointVariableAll}{0.0007}
\newcommand{\JDOffsetMinusTWOWASPTwelveJointVariableAll}{0.0010}
\newcommand{\JDOffsetTHREEWASPTwelveJointVariableAll}{5194.9356}
\newcommand{\JDOffsetPlusTHREEWASPTwelveJointVariableAll}{0.0007}
\newcommand{\JDOffsetMinusTHREEWASPTwelveJointVariableAll}{0.0010}

\newcommand{\ParamSixWASPTwelveJointVariableAll}{0.180}
\newcommand{\ParamSixPlusWASPTwelveJointVariableAll}{0.018}
\newcommand{\ParamSixMinusWASPTwelveJointVariableAll}{0.020}
\newcommand{\ParamSevenWASPTwelveJointVariableAll}{0.00115}
\newcommand{\ParamSevenPlusWASPTwelveJointVariableAll}{0.00026}
\newcommand{\ParamSevenMinusWASPTwelveJointVariableAll}{0.00024}
\newcommand{\ParamEightWASPTwelveJointVariableAll}{-0.003}
\newcommand{\ParamEightPlusWASPTwelveJointVariableAll}{0.001}
\newcommand{\ParamEightMinusWASPTwelveJointVariableAll}{0.001}
\newcommand{\ParamNineWASPTwelveJointVariableAll}{0.129}
\newcommand{\ParamNinePlusWASPTwelveJointVariableAll}{0.027}
\newcommand{\ParamNineMinusWASPTwelveJointVariableAll}{0.031}
\newcommand{\ParamTenWASPTwelveJointVariableAll}{0.00163}
\newcommand{\ParamTenPlusWASPTwelveJointVariableAll}{0.00031}
\newcommand{\ParamTenMinusWASPTwelveJointVariableAll}{0.00031}
\newcommand{\ParamElevenWASPTwelveJointVariableAll}{-0.015}
\newcommand{\ParamElevenPlusWASPTwelveJointVariableAll}{0.003}
\newcommand{\ParamElevenMinusWASPTwelveJointVariableAll}{0.003}
\newcommand{\TBrightSixWASPTwelveJointVariableAll}{2763}
\newcommand{\TBrightSixPlusWASPTwelveJointVariableAll}{80}
\newcommand{\TBrightSixMinusWASPTwelveJointVariableAll}{84}
\newcommand{\fReradiationSixWASPTwelveJointVariableAll}{0.353}
\newcommand{\fReradiationSixPlusWASPTwelveJointVariableAll}{0.043}
\newcommand{\fReradiationSixMinusWASPTwelveJointVariableAll}{0.041}
\newcommand{\TBrightNineWASPTwelveJointVariableAll}{2849}
\newcommand{\TBrightNinePlusWASPTwelveJointVariableAll}{138}
\newcommand{\TBrightNineMinusWASPTwelveJointVariableAll}{155}
\newcommand{\fReradiationNineWASPTwelveJointVariableAll}{0.399}
\newcommand{\fReradiationNinePlusWASPTwelveJointVariableAll}{0.083}
\newcommand{\fReradiationNineMinusWASPTwelveJointVariableAll}{0.080}

\newcommand{\PrecessionZeroLikely}{4508.9769}	
\newcommand{\PrecessionZeroPlus}{0.0001}	
\newcommand{\PrecessionZeroMinus}{0.0002}
\newcommand{\PrecessionOneLikely}{1.0914239}	
\newcommand{\PrecessionOnePlus}{0.0000004}	
\newcommand{\PrecessionOneMinus}{0.0000004}
\newcommand{\PrecessionTwoLikely}{0.00095}	
\newcommand{\PrecessionTwoPlus}{0.01365}	
\newcommand{\PrecessionTwoMinus}{0.00063}
\newcommand{\PrecessionThreeLikely}{-90.9}	
\newcommand{\PrecessionThreePlus}{184.5}	
\newcommand{\PrecessionThreeMinus}{4.5}
\newcommand{\PrecessionFourLikely}{0.003}	
\newcommand{\PrecessionFourPlus}{0.002}	
\newcommand{\PrecessionFourMinus}{0.001}
\newcommand{\PrecessionECosOmegaLikely}{-0.0000}
\newcommand{\PrecessionECosOmegaPlus}{0.0003}
\newcommand{\PrecessionECosOmegaMinus}{0.0008}
\newcommand{\PrecessionESinOmegaLikely}{-0.004}
\newcommand{\PrecessionESinOmegaPlus}{0.016}
\newcommand{\PrecessionESinOmegaMinus}{0.013}
\newcommand{\PrecessionBICLikely}{106.6}
\newcommand{\PrecessionChiSquaredLikely}{85.5}

\newcommand{\PrecessionZeroNoPrecessLikely}{4508.9769}	
\newcommand{\PrecessionZeroNoPrecessPlus}{0.0001}	
\newcommand{\PrecessionZeroNoPrecessMinus}{0.0001}
\newcommand{\PrecessionOneNoPrecessLikely}{1.0914238}	
\newcommand{\PrecessionOneNoPrecessPlus}{0.0000003}	
\newcommand{\PrecessionOneNoPrecessMinus}{0.0000002}
\newcommand{\PrecessionTwoNoPrecessLikely}{0.00089}	
\newcommand{\PrecessionTwoNoPrecessPlus}{0.02568}	
\newcommand{\PrecessionTwoNoPrecessMinus}{0.00057}
\newcommand{\PrecessionThreeNoPrecessLikely}{-89.9}	
\newcommand{\PrecessionThreeNoPrecessPlus}{1.8}	
\newcommand{\PrecessionThreeNoPrecessMinus}{2.7}

\newcommand{\PrecessionECosOmegaNoPrecessLikely}{0.0000}
\newcommand{\PrecessionECosOmegaNoPrecessPlus}{0.0004}
\newcommand{\PrecessionECosOmegaNoPrecessMinus}{0.0005}
\newcommand{\PrecessionESinOmegaNoPrecessLikely}{-0.015}
\newcommand{\PrecessionESinOmegaNoPrecessPlus}{0.015}
\newcommand{\PrecessionESinOmegaNoPrecessMinus}{0.018}
\newcommand{\PrecessionBICNoPrecessLikely}{110.1}
\newcommand{\PrecessionChiSquaredNoPrecessLikely}{93.2}

\newcommand{\PrecessionZeroNoPrecessNoLopezLikely}{4508.9769}	
\newcommand{\PrecessionZeroNoPrecessNoLopezPlus}{0.0002}	
\newcommand{\PrecessionZeroNoPrecessNoLopezMinus}{0.0001}
\newcommand{\PrecessionOneNoPrecessNoLopezLikely}{1.0914239}	
\newcommand{\PrecessionOneNoPrecessNoLopezPlus}{0.0000002}	
\newcommand{\PrecessionOneNoPrecessNoLopezMinus}{0.0000004}
\newcommand{\PrecessionTwoNoPrecessNoLopezLikely}{0.00087}	
\newcommand{\PrecessionTwoNoPrecessNoLopezPlus}{0.02494}	
\newcommand{\PrecessionTwoNoPrecessNoLopezMinus}{0.00058}
\newcommand{\PrecessionThreeNoPrecessNoLopezLikely}{-90.0}	
\newcommand{\PrecessionThreeNoPrecessNoLopezPlus}{180.9}	
\newcommand{\PrecessionThreeNoPrecessNoLopezMinus}{2.7}

\newcommand{\PrecessionECosOmegaNoPrecessNoLopezLikely}{-0.0001}
\newcommand{\PrecessionECosOmegaNoPrecessNoLopezPlus}{0.0005}
\newcommand{\PrecessionECosOmegaNoPrecessNoLopezMinus}{0.0005}
\newcommand{\PrecessionESinOmegaNoPrecessNoLopezLikely}{-0.015}
\newcommand{\PrecessionESinOmegaNoPrecessNoLopezPlus}{0.017}
\newcommand{\PrecessionESinOmegaNoPrecessNoLopezMinus}{0.019}
\newcommand{\PrecessionBICNoPrecessNoLopezLikely}{108.2}
\newcommand{\PrecessionChiSquaredNoPrecessNoLopezLikely}{91.4}

\newcommand{\RWideWASPTwelve}{1.9}
\newcommand{\RWideJupiter}{3.3}
\newcommand{\BetaJ}{1.7}
\newcommand{\BetaH}{1.3}
\newcommand{\BetaKs}{1.1}

\newcommand{\fReradiationWASPTwelveALL}{0.441}
\newcommand{\fReradiationPlusWASPTwelveALL}{0.024}
\newcommand{\fReradiationMinusWASPTwelveALL}{0.023}
\newcommand{\NightsidePercentage}{11}
\newcommand{\DaysidePercentage}{89}
\newcommand{\BolometicFluxTotal}{1.25}
\newcommand{\BolometricFluxDayside}{1.12}

\newcommand{\BlackbodyOneChi}{34}
\newcommand{\BlackbodyTwoChi}{5}
\newcommand{\FortneyOneChi}{83}
\newcommand{\FortneyTwoChi}{72}
\newcommand{\FortneyThreeChi}{20}
\newcommand{\FortneyFourChi}{20}
\newcommand{\XPointXXKs}{10.0$\times$10$^{-3}$}
\newcommand{\YPointYYKs}{0.86$\times$10$^{-3}$}
\newcommand{\XPointXXH}{3.8$\times$10$^{-3}$}
\newcommand{\YPointYYH}{0.90$\times$10$^{-3}$}
\newcommand{\XPointXXJ}{1.93$\times$10$^{-3}$}
\newcommand{\YPointYYJ}{0.69$\times$10$^{-3}$}

\newcommand{\HowManyKs}{17}
\newcommand{\HowManyH}{7}
\newcommand{\HowManyJ}{22}

\begin{document}

\title{Near-infrared thermal emission from WASP-12b: 
detections of the secondary eclipse in Ks, H \& J\altaffilmark{*}}

\author{Bryce Croll\altaffilmark{1},
David Lafreniere\altaffilmark{2},
Loic Albert\altaffilmark{3},
Ray Jayawardhana\altaffilmark{1},
Jonathan J. Fortney\altaffilmark{4},
Norman Murray\altaffilmark{5,6}
}

\altaffiltext{1}{Department of Astronomy and Astrophysics, University of Toronto, 50 St. George Street, Toronto, ON 
M5S 3H4, Canada;
croll@astro.utoronto.ca}

\altaffiltext{2}{D\'epartement de physique, Universit\'e de Montr\'eal, C.P.
6128 Succ. Centre-Ville, Montr\'eal, QC, H3C 3J7, Canada}

\altaffiltext{3}{Canada-France-Hawaii Telescope Corporation, 65-1238 Mamalahoa Highway,
Kamuela, HI 96743.}

\altaffiltext{4}{Department of Astronomy and Astrophysics, University of California, Santa Cruz, CA, 95064}

\altaffiltext{5}{Canadian Institute for Theoretical Astrophysics, 60 St. George Street, University of Toronto, Toronto ON M5S 3H8, Canada}

\altaffiltext{6}{Canada Research Chair in Astrophysics}

\altaffiltext{*}{Based on observations obtained with WIRCam, a joint project of CFHT, Taiwan, Korea, Canada, France, at the Canada-France-Hawaii Telescope (CFHT) which is operated by the National Research Council (NRC) of Canada, the Institute National des Sciences de l'Univers of the Centre National de la Recherche Scientifique of France, and the University of Hawaii.}

\begin{abstract}
We present Ks, H \& J-band photometry of the very highly irradiated hot Jupiter WASP-12b 
using the Wide-field Infrared Camera on the Canada-France-Hawaii telescope. 
Our photometry brackets the secondary eclipse of WASP-12b in the Ks and H-bands,
and in J-band starts in mid-eclipse
and continues until well after the end of the eclipse.
We detect its thermal emission in all three near-infrared bands.
Our secondary eclipse depths are 
\FpOverFStarPercentAbstractWASPTwelveJointAll$^{+\FpOverFStarPercentAbstractPlusWASPTwelveJointAll}_{-\FpOverFStarPercentAbstractMinusWASPTwelveJointAll}$\% in Ks-band (\XSigmaWASPTwelveKs $\sigma$),
\ParamSixWASPTwelveJointAll$^{+\ParamSixPlusWASPTwelveJointAll}_{-\ParamSixMinusWASPTwelveJointAll}$\% in H-band (\XSigmaSixWASPTwelveJointAll $\sigma$) and
\ParamNineWASPTwelveJointAll$^{+\ParamNinePlusWASPTwelveJointAll}_{-\ParamNineMinusWASPTwelveJointAll}$\% in J-band (\XSigmaNineWASPTwelveJointAll $\sigma$).
All three secondary eclipses are best-fit with a consistent phase, $\phi$, that is
compatible with a 
circular orbit: $\phi$=\PhaseAbstractWASPTwelveJointAll$^{+\PhaseAbstractPlusWASPTwelveJointAll}_{-\PhaseAbstractMinusWASPTwelveJointAll}$.
The limits on the eccentricity, $e$,
and argument of periastron, $\omega$, of 
this planet from our photometry alone are thus
$|$$e$$\cos$$\omega$$|$ $<$ \ECosOmegaAbsoluteThreeSigmaLimitWASPTwelveJointAll.
By combining our secondary eclipse times with others published in the literature,
as well as the radial velocity and transit timing data for this system, 
we show that there is no evidence that WASP-12b is precessing at a detectable rate, and show that its orbital 
eccentricity is likely zero.
Our thermal emission measurements also allow us to constrain the characteristics 
of the planet's atmosphere; 
%ATMOSPHERIC_CHANGE
%our Ks-band and H-band eclipse depths argue strongly in favour of inefficient day to nightside
%redistribution of heat and a low Bond albedo for this very highly irradiated hot Jupiter.
%Our J-band eclipse is shallower than expected for a planet that redistributes heat inefficiently to 
%the nightside of the planet
%and hints at the possibility of a temperature inversion
%deep in the atmosphere of WASP-12b; the most likely explanation is that the high pressure, deep atmospheric layer
%probed by these observations is more homogenized than higher altitude layers.
%ATMOSPHERIC_CHANGE
our Ks-band eclipse depth argues strongly in favour of inefficient day to nightside
redistribution of heat and a low Bond albedo for this very highly irradiated hot Jupiter.
The J and H-band brightness temperatures are slightly cooler than the Ks-band brightness temperature, 
and thus hint at the possibility of a modest temperature inversion
deep in the atmosphere of WASP-12b; the high pressure, deep atmospheric layers
probed by our J and H-band observations are likely more homogenized
than the higher altitude layer probed by our Ks-band observations.
Lastly,
our best-fit Ks-band eclipse has a marginally 
longer duration than would otherwise be expected;
this may be tentative evidence for material being tidally 
stripped from the planet -- 
as was predicted for this system by Li \& collaborators, 
and for which observational confirmation was recently arguably provided by Fossati \& collaborators.
\end{abstract}

\keywords{planetary systems . stars: individual: WASP-12 . techniques: photometric-- eclipses -- infrared: planetary systems}

\section{Introduction}

Multiwavelength constraints on the thermal emission of hot Jupiters are crucial to precisely defining
the spectral energy distributions of these planets and understanding their energy budgets.
Interestingly most hot Jupiter
thermal emission detections to date have not been at the blackbody peaks of these planets, but at longer
wavelengths with the Spitzer Space Telescope
($\lambda$ $>$ 3 $\mu m$; e.g. \citealt{Charbonneau05,Deming05}).
Probing shorter near-infrared wavelengths at  
the blackbody peaks of these planets has only recently been proven feasible first through space-based
observations with the Hubble Space Telescope (HST; \citealt{Swain09}),
and then from the ground (e.g. \citealt{deMooij09,Sing09,Gillon09}). 
Our program to detect near-infrared thermal emission from the hottest of the 
hot Jupiters has also been successful using the 
Wide-field Infrared Camera (WIRCam) on the Canada-France-Hawaii Telescope (CFHT) to detect the Ks-band thermal
emission of: TrES-2b \citep{CrollTrESTwo}, TrES-3b including an H-band upper-limit \citep{CrollTrESThree},
and two eclipses of WASP-3b, including a limit on its temporal variability \citep{CrollWASPThree}.

In the near-infrared multiple band detections have only been performed on a handful of occasions;
such multiple-band detections were performed in narrow wavelength regimes from space via spectroscopy with
HST for HD 209458 and HD 189733 \citep{Swain09,Swain09b}, and arguably
recently from the ground for HD 189733 using the Infrared Telescope Facility \citep{Swain10}, as well as from the ground
using the Very Large Telescope in the H \& K-bands for the highly irradiated
hot Jupiter WASP-19b \citep{Anderson10,Gibson10}.
Multiple band detections in
the near-infrared are therefore rare compared to the frequent
multiple-band detections at longer wavelengths using the IRAC \citep{Fazio04}, IRS \citep{Houck04}, or MIPS \citep{Rieke04} instruments
on the Spitzer Space Telescope. 
Multiwavelength thermal emission measurements with Spitzer have revealed a wealth of information, including that the most highly irradiated exoplanets
seem to harbour hot stratospheres and temperature inversions \citep{Knutson08,Charbonneau08,Machalek08,Knutson08b}.
One could imagine that obtaining multiwavelength constraints on a planet's thermal emission in the near-infrared could be equally informative.
Furthermore the near-infrared is also an ideal place to directly constrain these planets' pressure-temperature profiles at depth, dayside bolometric luminosities
and the fraction of the incident stellar radiation that is transported from the tidally locked 
day to nightsides deep in these planets' atmospheres \citep{Barman08}.

Here we continue our program using WIRCam on CFHT to detect thermal emission from 
some of the hottest of the hot Jupiters. Our target was
the highly irradiated hot Jupiter WASP-12b.
The discovery of the inflated, transiting exoplanet WASP-12b
was of immediate interest to those attempting to measure the loss in flux 
during the secondary eclipses of hot Jupiters in the near-infrared - this was because WASP-12b circles
a late F-type star with a period of only $\sim$26 hours \citep{Hebb09}.
It is thus exposed to extremely
high stellar insolation, with an incident flux of $\sim$$9$$\times$$10^{9}$ $erg$$s^{-1}$$cm^{-2}$.
The planet is also one of the most inflated hot Jupiters, with a radius of $R_{P}$$\sim$1.8$R_{J}$ and 
a favourable planet-to-star radius ratio ($R_{P}/R_{*}$$\sim$0.12; \citealt{Hebb09}). It should
be heated to an equilibrium temperature of over $\sim$2500 $K$
assuming isotropic reradiation and a zero Bond albedo\footnote{The Bond albedo is the fraction of the bolometric flux reflected from the planet compared to the incident bolometric radiation.}.
For these reasons it was predicted 
to display near-infrared thermal emission on the order of 0.1-0.3\% of the stellar flux in the J, H \& Ks near-infrared
bands, assuming isotropic reradiation and a zero Bond albedo.
\citet{LopezMorales10} have already reported a detection of the secondary eclipse of WASP-12b in z'-band 
(0.9 $\mu m$), and more recently \citet{Campo10} have presented detections of two eclipses in the four
IRAC channels for WASP-12b. \citet{Campo10}, however, did not report the eclipse depths for WASP-12b, and 
for reasons discussed below the \citet{LopezMorales10} detection has recently been called into question. 
Thus the atmospheric characteristics of WASP-12b remain largely unconstrained.

	In addition to receiving extremely high stellar insolation, WASP-12b is intriguing because the combination of
its close proximity
to its star and its putative original eccentricity ($e$=0.049$\pm$0.015; \citealt{Hebb09}) suggests that it could be 
precessing at a rate that is detectable with current instruments.
Such a putative precession signal was recently claimed by \citet{Campo10}.
Although the IRAC eclipses reported by \citet{Campo10} suggest an $e$cos$\omega$ constraint similar to that expected for a circular orbit 
($e$cos$\omega$$=$-0.0054$\pm$0.0030), \citet{LopezMorales10} had earlier reported an
%proofdiff
%considerably offset from a circular orbit ($e$cos$\omega$$=$0.0156$\pm$0.0035). While at first glance the two measurements
eclipse detection that was 
considerably offset from a circular orbit ($|$$e$cos$\omega$$|$$=$0.016$^{+0.011}_{-0.009}$).
While at first glance the two measurements
are inconsistent, if the planet precesses this is not the case. By combining their secondary eclipses with those of
\citet{LopezMorales10},
together with the original radial velocity data for the system \citep{Hebb09}, as well as a series of transit-time measurements
from the original detection paper and ground-based amateurs (from the Exoplanet Transit
Database; \citealt{Poddany10}), \citet{Campo10} show that a precessing orbital 
model best-fits the data with 2$\sigma$ confidence. 
The authors caution that this detection is heavily dependent on the secondary eclipse offset
reported by \citet{LopezMorales10}. 
Even more recently, 
radial velocity observations of WASP-12b have suggested that the eccentricity of WASP-12b is small
($e$=0.017$^{+0.015}_{-0.010}$; \citealt{Husnoo10}) and likely zero, constraining the
\citet{Campo10} precession signal and calling into question
the \citet{LopezMorales10} eclipse detection.
Nevertheless, the best-fit eccentricity of WASP-12b remains non-zero, and thus this planet
could be precessing at a much slower rate than 
\citet{Campo10} claim. The definitive nail in the coffin on the claim that WASP-12b is precessing
at a detectable rate, will thus only result from further detections
of this planet's secondary eclipse well seperated in time from the original eclipse detections. 

 Also, recently preliminary evidence was presented that material from WASP-12b may be being tidally
stripped from the planet and may possibly form a circumstellar disk 
in this system. \citet{Li10} predicted this system may have such a disk from material overfilling the Roche lobe 
of WASP-12b, because 
WASP-12b's observed radius in the optical ($R_p$$\sim$1.79 $R_{J}$; \citealt{Hebb09}) 
is already close to its 2.36 $R_{J}$ Roche lobe radius (as quoted in \citealt{Fossati10}).
That WASP-12b may exhibit material overfilling it Roche lobe and a circumstellar disk from this material
has recently received possible confirmation from HST Cosmic Origins Spectrograph (COS)
observations of this system.
From these observations \citet{Fossati10} find increased transit depths in the ultraviolet
when compared to the optical, 
indicative of material surrounding WASP-12b overfilling its Roche lobe and blocking out a larger fraction
of the stellar flux at these wavelengths. In addition they observe an early ingress
of the transit of WASP-12b in their near ultraviolet data;
\citet{Fossati10} intrepret this early ingress as a putative sign of 
previously stripped material from WASP-12b forming a circumstellar disk.
These putative signs of a disk are interesting to observers in the near-infrared, specifically the K-band,
as \citet{Li10} predicted that such a disk in this system may exhibit CO emission
as bright as 10 mJy at 2.292 $\mu m$. WASP-12 does not, however, display a significant near-infrared excess \citep{Fossati10b}.

	Here we present detections of WASP-12b's thermal emission in the Ks (\XSigmaWASPTwelveJointAll$\sigma$), H (\XSigmaSixWASPTwelveJointAll$\sigma$)
and J-bands (\XSigmaNineWASPTwelveJointAll$\sigma$). Our J-band detection is the first thermal emission
measurement in  this band.
Our photometry favours a circular orbit for WASP-12b 
($e$$\cos$$\omega$$=$\ECosOmegaWASPTwelveJointAll$^{+\ECosOmegaPlusWASPTwelveJointAll}_{-\ECosOmegaMinusWASPTwelveJointAll}$). 
By combining our secondary eclipse times with those of \citet{LopezMorales10} and \citet{Campo10}, as well as the radial velocity data of \citet{Hebb09}
and \citet{Husnoo10}, and all the transit-time data for the system, we are able to show that
not only is there no evidence to date that WASP-12b is precessing at a detectable rate,
but also that the orbit of WASP-12b is likely circular.
Our analysis also allows us to constrain the characteristics of the atmospere of WASP-12b;
%our H \& Ks-band eclipse depths argue in favour of inefficient redistribution of heat from the day to nightside,
%while our J-band observations seem to be probing a deeper, higher pressure atmospheric
%layer that is more homogenized.
%ATMOSPHERIC_CHANGE
our Ks-band eclipse depth argues in favour of inefficient redistribution of heat from the day to nightside,
while our J and H-band observations seem to be probing deeper, higher pressure atmospheric
layers that are slightly more homogenized.
We also show that our Ks-band photometry may feature a longer than expected eclipse duration that could
arguably be interpreted as evidence for material streaming from the planet or a circumstellar disk in this system.

\section{Observations and Data Reduction}
\label{SecReduction}

\begin{figure}
\centering
%EMULATEAPJCHANGE
\includegraphics[scale=0.40,angle=90]{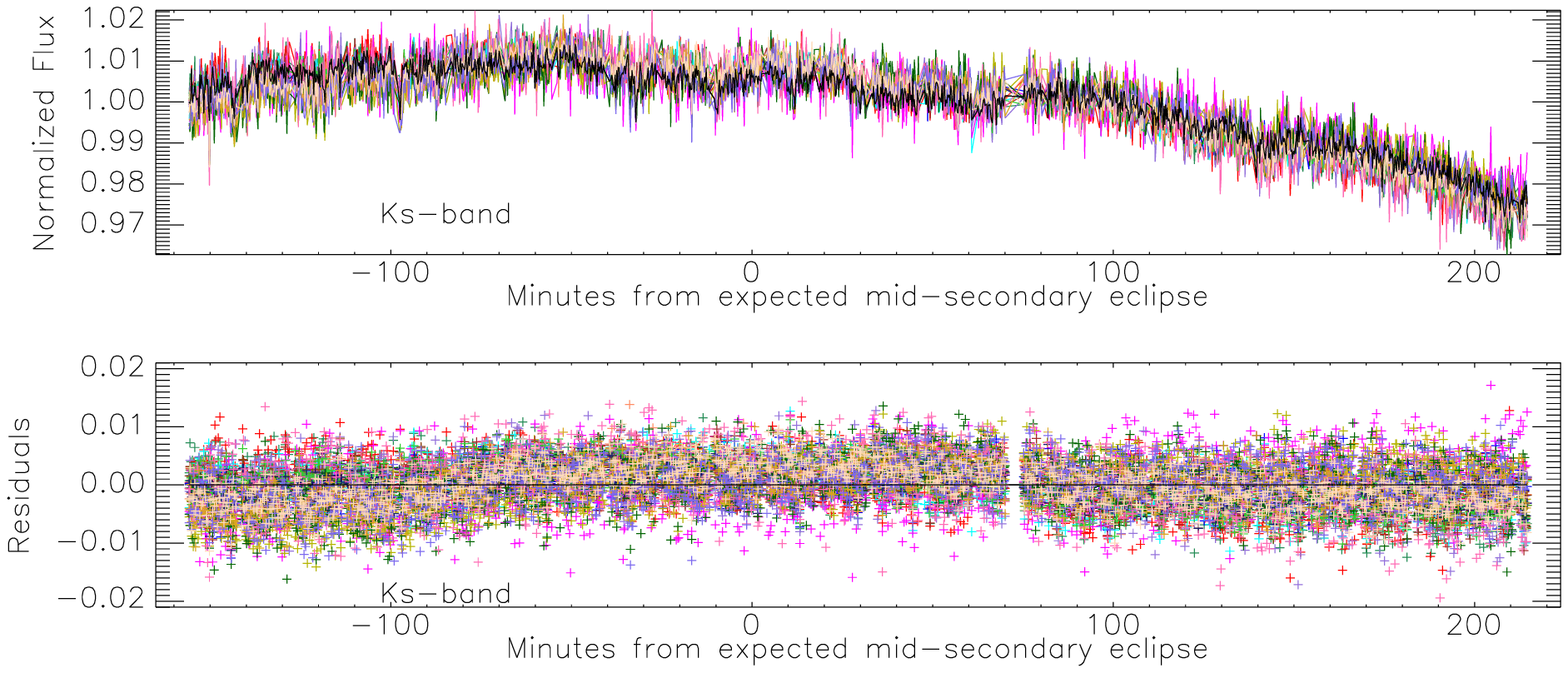}
\includegraphics[scale=0.40,angle=90]{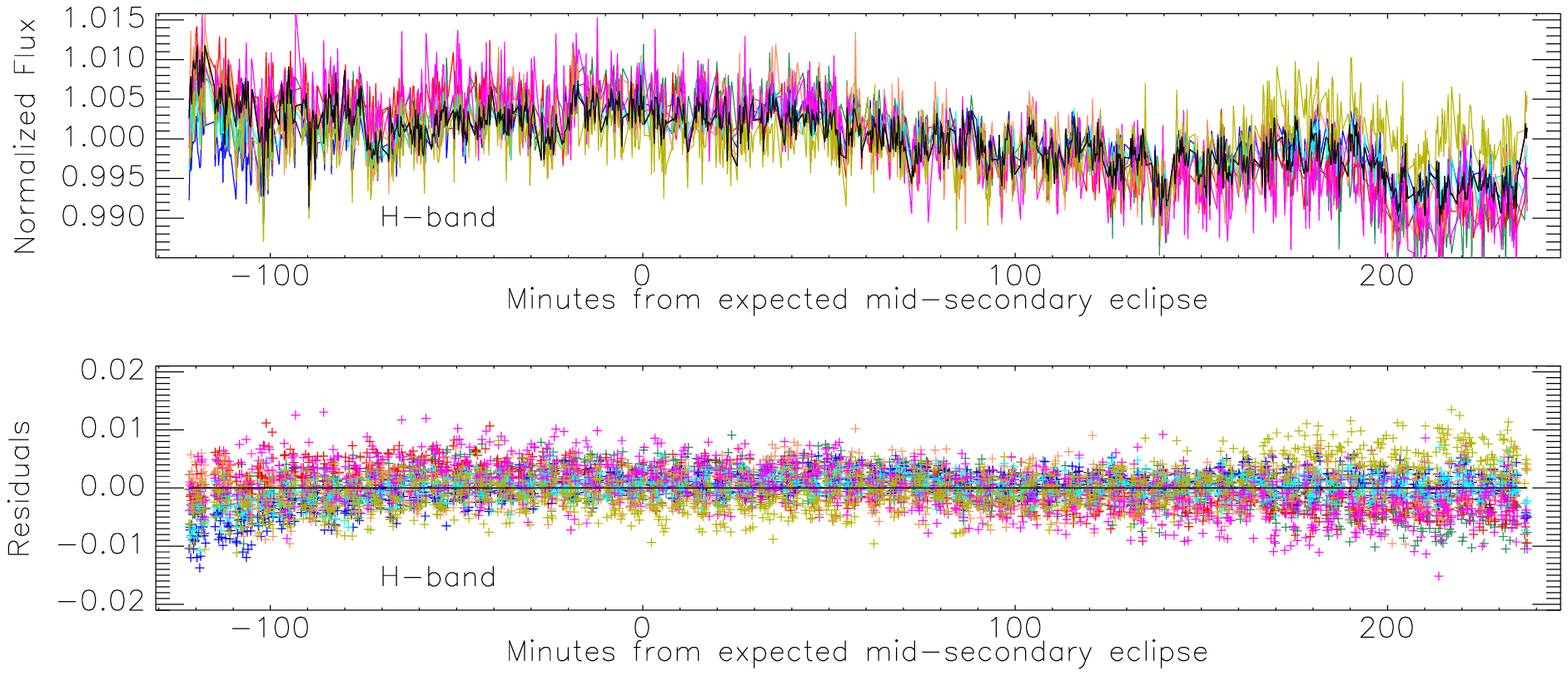}
\includegraphics[scale=0.40,angle=90]{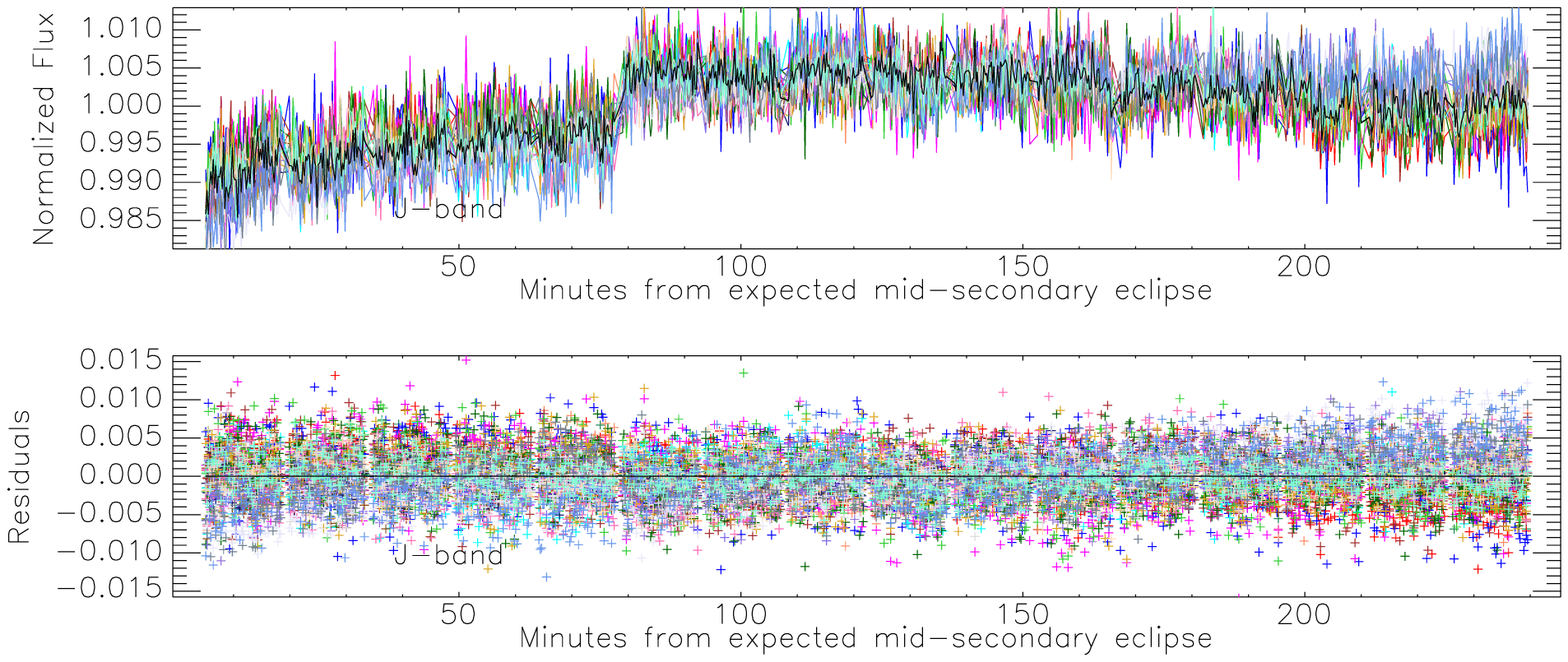}
\caption{	The normalized flux from our target star and reference stars for our Ks-band photometry (top two panels),
		our H-band photometry (middle two panels), and our J-band photometry (bottom panels).
		For each set of panels the top panel displays the flux 
		from the target star (black) and the reference stars (various colours)
		that are used to calibrate the flux of WASP-12b in the various sets of photometry.
		The bottom panels in each set of panels displays the 
		residuals from the normalized flux of the target star corrected by the normalized flux of the reference stars.
	}
\label{FigWASP12RefStars}
\end{figure}

\begin{figure}
\centering
\includegraphics[scale=0.45, angle = 270]{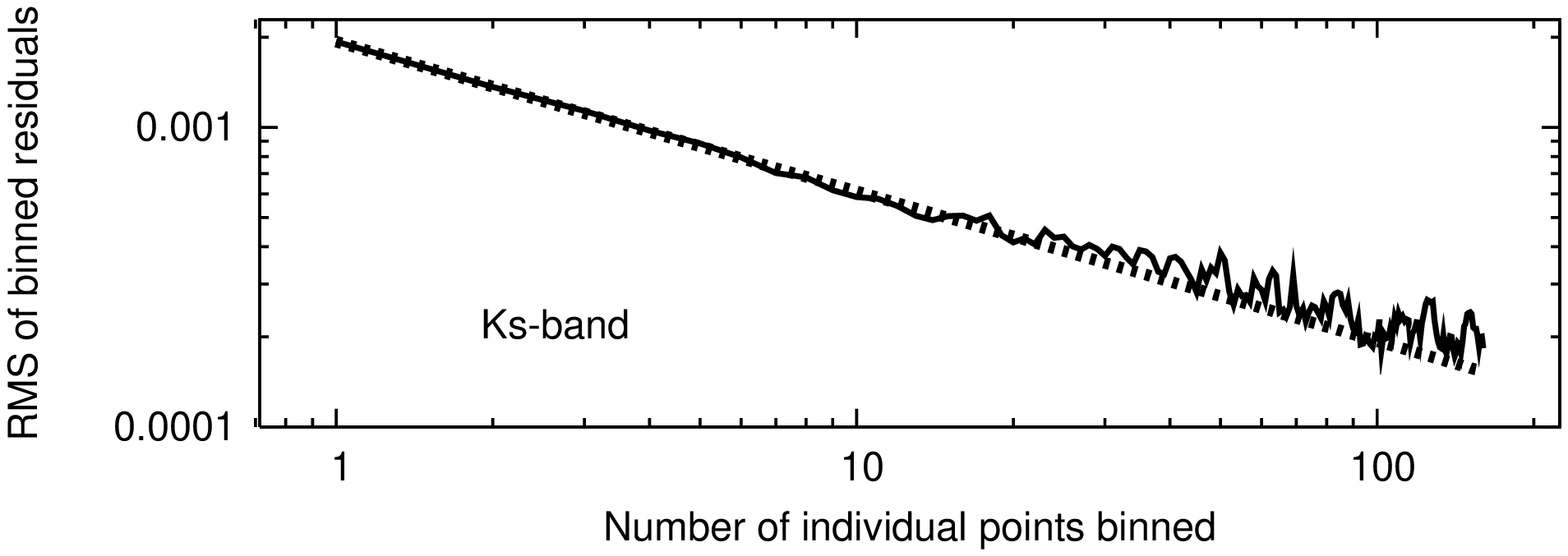}
\includegraphics[scale=0.45, angle = 270]{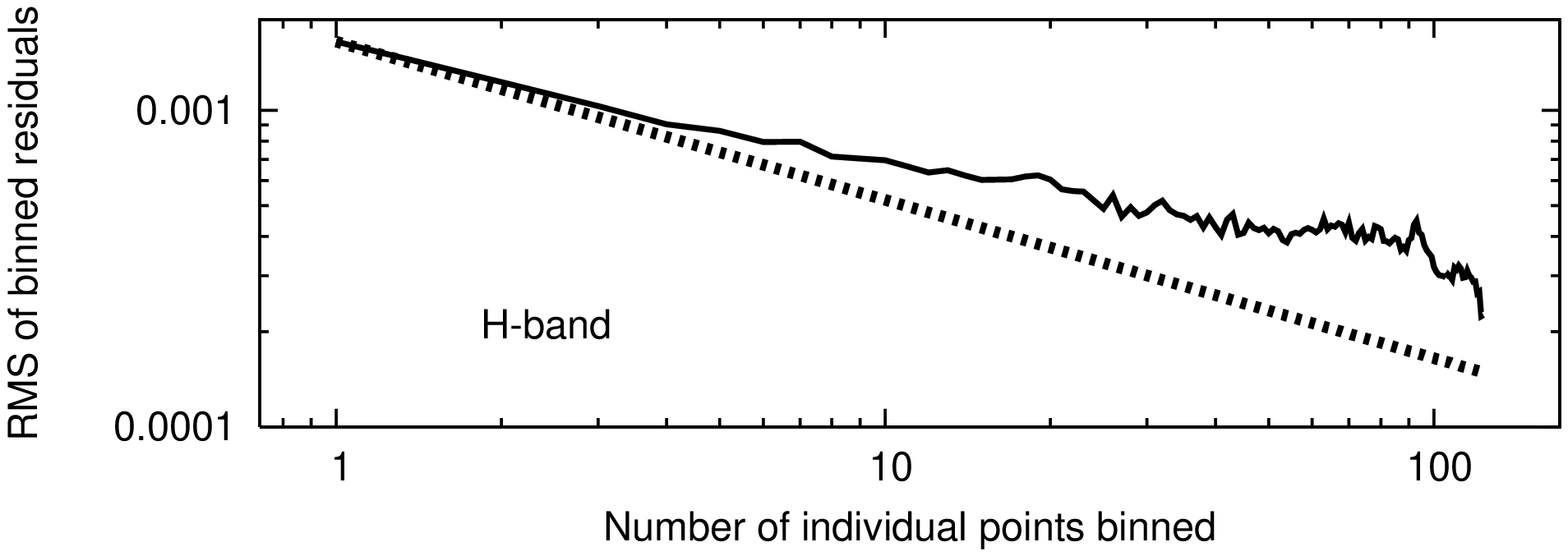}
\includegraphics[scale=0.45, angle = 270]{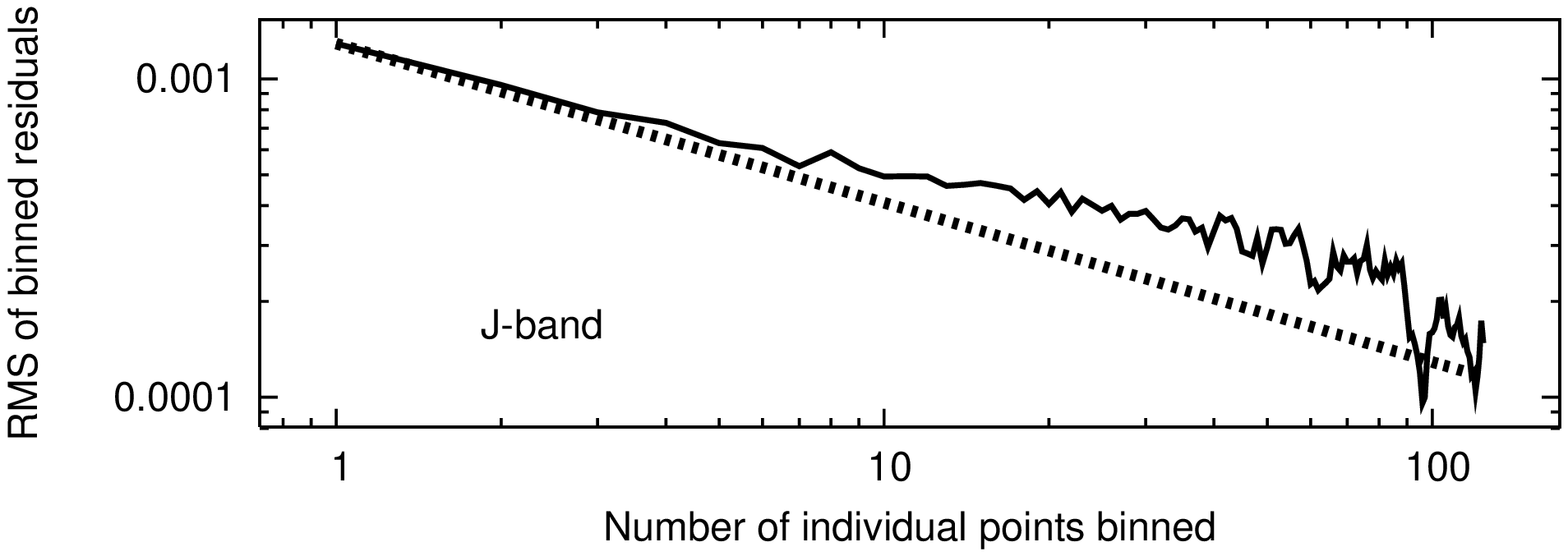}
\caption{	The root-mean-square of our out-of-eclipse photometry (solid line) 
		following the corrections documented in $\S$\ref{SecReduction}
		for our Ks-band photometry (top), our H-band photometry (middle), and our J-band photometry (bottom).
		The dashed line in each panel displays the
		one over the square-root of the bin-size expectation for gaussian noise.
		}
\label{FigPoisson}
\end{figure}

We obtained observations with WIRCam on CFHT of WASP-12 ($J$$\sim$10.48, $H$$\sim$10.23, $K$$\sim$10.19)
on 2009 December 26, 27 and 28
in the J, H \& Ks-bands respectively.
Our J-band observations on Dec. 26 lasted for 3.9 hours and started in mid-eclipse and persisted
for 2.2 hours after the end of eclipse.
Our observations on Dec. 27 and 28 lasted 
for 6.0 hours in H-band and 6.2 hours in Ks-band, respectively, evenly bracketing the
predicted secondary eclipse of WASP-12b.
Numerous reference stars were also observed in the 21$\times$21 arcmin
field-of-view of WIRCam. The telescope was defocused for our various observations to approximately
1.5mm (J-band), 1.8mm (H-band), and 2.0mm (Ks-band), resulting
in the flux of our target star being spread over a ring $\sim$19, $\sim$23, and $\sim$26  pixels in diameter 
(6, 7 and 8\arcsec) on our array.
For each observation, as the telescope temperature changed over the course of the night, we
used the focus stage model and kept the defocus amount constant, thus achieved a
stable PSF over the entire observation set.
We used ``Staring Mode'' for our J and Ks-band observations where we do not dither for the duration of our observations; for the
H-band eclipse the queue observations mistakenly used micro-dithering which featured small 0.5 pixel shifts between
consecutive exposures.
The exposure times for our J, H \& Ks-band observations were 5-seconds.
The effective duty cycle after accounting for readout and for saving exposures was 34\%. 

For both observations the data was reduced and aperture photometry was performed on our target
star and our reference stars as discussed in \citet{CrollTrESThree}
(with the details provided in \citealt{CrollTrESTwo}).
We used an aperture with a radius of 17 pixels for our Ks-band photometry, and 16.5 pixels for our H and J-band photometry.
We used an annulus to define the sky with an inner
radius of 22, and an outer radius of 34 pixels for all our photometry.
We ensured that these choices of aperture were optimal
by testing smaller and larger aperture sizes in increments of 0.5 pixels and ensuring 
these choices displayed the smallest root-mean-square (RMS) outside of occultation and the least
time-correlated red-noise.

Following our aperture photometry we correct the flux of our target star with a number of nearby reference stars
as discussed in \citet{CrollTrESTwo}.
We used \HowManyJ, \HowManyH, and \HowManyKs \ reference stars to correct our J, H and Ks-band eclipse photometry, respectively.
The normalized flux of WASP-12 and the various reference stars that are used
to correct the flux of our target star
are displayed in Figure \ref{FigWASP12RefStars}.
For our Ks-band photometry we corrected our photometry for a small trend 
with the x, and y pixel position of the target star on the chip \footnote{The correction is described in \citet{CrollTrESTwo}}. We didn't notice such trends in our H and J-band photometry.

For our H-band photometry the airmass, $X$,
was high at the start of the observations (X$\sim$1.9), and fell to $X$$\sim$1.2 by mid-eclipse.
We noticed a downward trend in our H-band photometry following the correction with
nearby reference stars that appeared to be correlated with airmass.
We found this effect
was reduced, but not removed, for our H-band photometry
by correcting the flux of WASP-12 with reference stars solely on the same WIRCam chip as WASP-12; this downward trend 
in flux of our target star compared to the reference stars is still
apparent at the start of our H-band photometry.
To reduce the impact of these systematic data we scale-up the errors of the first $\sim$25 minutes of data for our H-band photometry
by a factor of 1.3.

The root-mean-square (RMS) of our photometry per minute following the above corrections improved from
\XPointXXKs \ to \YPointYYKs \ in Ks-band, \XPointXXH \ to \YPointYYH \ in H-band and \XPointXXJ \ to \YPointYYJ \ in J-band.
To evaluate the impact of systematics and the presence of red-noise in our photometry we bin our data and compare 
the out-of-eclipse photometric precision to the gaussian noise expectation of one over the square-root of the bin-size (Figure \ref{FigPoisson}).
The Ks-band data bins down very close to the gaussian noise limit, while
the J-band data bins down marginally above this limit;
the H-band data is worse, possibly due to the 
systematics introduced by the micro-dithering.
To ensure we do not underestimate the uncertainties in our model parameters we 
employ the \citet{Winn08} method to account for time-correlated red-noise in our photometry.
We scale-up the uncertainties
on our individual data-points by a factor $\beta$; $\beta$ is equal to the factor that the 
binned out-of-eclipse RMS scales above the gaussian noise expectation in the absence of red-noise. 
We use a binning time of $\sim$12 minutes.
For our H-band photometry we exclude the first 15 minutes of data
from this calculation, due to the obvious systematic that we believe to be correlated with airmass that
does not appear to affect the rest of the photometry.
For our photometry
$\beta$ is equal to \BetaJ, \BetaH \ and \BetaKs \ for our
J, H \& Ks-band data, respectively.
We note that our observations are still well above the predicted photon noise RMS
limit of 2.7$\times$10$^{-4}$, 2.2$\times$10$^{-4}$
and 3.1$\times$10$^{-4}$ per minute in the J, H \& Ks-bands, respectively. 

\section{Analysis}

%EMULATEAPJCHANGE
\begin{deluxetable*}{ccccccc}
%\begin{deluxetable}{ccccccc}
\setlength{\tabcolsep}{0.01in} 
\tablecaption{Best-fit secondary eclipse parameters}
\tabletypesize{\scriptsize}
\tablehead{
\colhead{Parameter} 	& \colhead{Ks-band} 		& \colhead{H-band}	& \colhead{J-band}	& \colhead{Joint}	& \colhead{Ks-band MCMC}	& \colhead{Joint MCMC}	\\	
\colhead{}		& \colhead{MCMC}		& \colhead{MCMC}	& \colhead{ MCMC}	& \colhead{MCMC}	& \colhead{variable eclipse} 	& \colhead{variable eclipse} \\
\colhead{}		& \colhead{Solution}		& \colhead{Solution}	& \colhead{Solution}	& \colhead{Solution}	& \colhead{duration solution} 	& \colhead{duration solution} \\
}
\startdata
reduced $\chi^{2}$			&	\ChiWASPTwelveKs$^{+\ChiPlusWASPTwelveKs}_{-\ChiMinusWASPTwelveKs}$										& \ChiWASPTwelveH$^{+\ChiPlusWASPTwelveH}_{-\ChiMinusWASPTwelveH}$										& \ChiWASPTwelveJ$^{+\ChiPlusWASPTwelveJ}_{-\ChiMinusWASPTwelveJ}$										& \ChiWASPTwelveJointAll$^{+\ChiPlusWASPTwelveJointAll}_{-\ChiMinusWASPTwelveJointAll}$ 										& \ChiWASPTwelveVariableKs$^{+\ChiPlusWASPTwelveVariableKs}_{-\ChiMinusWASPTwelveVariableKs}$										& \ChiWASPTwelveJointVariableAll$^{+\ChiPlusWASPTwelveJointVariableAll}_{-\ChiMinusWASPTwelveJointVariableAll}$ \\
$\Delta F_{Ks}$				&	\FpOverFStarPercentAbstractWASPTwelveKs$^{+\FpOverFStarPercentAbstractPlusWASPTwelveKs}_{-\FpOverFStarPercentAbstractMinusWASPTwelveKs}$\%	& n/a																		& n/a																		& \FpOverFStarPercentAbstractWASPTwelveJointAll$^{+\FpOverFStarPercentAbstractPlusWASPTwelveJointAll}_{-\FpOverFStarPercentAbstractMinusWASPTwelveJointAll}\% $		& \FpOverFStarPercentAbstractWASPTwelveVariableKs$^{+\FpOverFStarPercentAbstractPlusWASPTwelveVariableKs}_{-\FpOverFStarPercentAbstractMinusWASPTwelveVariableKs}$\%	& \FpOverFStarPercentAbstractWASPTwelveJointVariableAll$^{+\FpOverFStarPercentAbstractPlusWASPTwelveJointVariableAll}_{-\FpOverFStarPercentAbstractMinusWASPTwelveJointVariableAll}$\% \\
$\Delta F_H$				&	n/a																		& \FpOverFStarPercentAbstractWASPTwelveH$^{+\FpOverFStarPercentAbstractPlusWASPTwelveH}_{-\FpOverFStarPercentAbstractMinusWASPTwelveH}$\%	& n/a																		& \ParamSixWASPTwelveJointAll$^{+\ParamSixPlusWASPTwelveJointAll}_{-\ParamSixMinusWASPTwelveJointAll}$\% 								& n/a																					& \ParamSixWASPTwelveJointVariableAll$^{+\ParamSixPlusWASPTwelveJointVariableAll}_{-\ParamSixMinusWASPTwelveJointVariableAll}$\% 	\\
$\Delta F_J$				&	n/a																		& n/a																		& \FpOverFStarPercentAbstractWASPTwelveJ$^{+\FpOverFStarPercentAbstractPlusWASPTwelveJ}_{-\FpOverFStarPercentAbstractMinusWASPTwelveKs}$\%	& \ParamNineWASPTwelveJointAll$^{+\ParamNinePlusWASPTwelveJointAll}_{-\ParamNineMinusWASPTwelveJointAll}$\% 								& n/a																					& \ParamNineWASPTwelveJointVariableAll$^{+\ParamNinePlusWASPTwelveJointVariableAll}_{-\ParamNineMinusWASPTwelveJointVariableAll}$\% \\
$t_{offset}$ ($min$)\tablenotemark{a}	&	\TOffsetWASPTwelveKs$^{+\TOffsetPlusWASPTwelveKs}_{-\TOffsetMinusWASPTwelveKs}$									& \TOffsetWASPTwelveH$^{+\TOffsetPlusWASPTwelveH}_{-\TOffsetMinusWASPTwelveH}$									& \TOffsetWASPTwelveJ$^{+\TOffsetPlusWASPTwelveJ}_{-\TOffsetMinusWASPTwelveJ}$									& \TOffsetWASPTwelveJointAll$^{+\TOffsetPlusWASPTwelveJointAll}_{-\TOffsetMinusWASPTwelveJointAll}$ 									& \TOffsetWASPTwelveVariableKs$^{+\TOffsetPlusWASPTwelveVariableKs}_{-\TOffsetMinusWASPTwelveVariableKs}$								& \TOffsetWASPTwelveJointVariableAll$^{+\TOffsetPlusWASPTwelveJointVariableAll}_{-\TOffsetMinusWASPTwelveJointVariableAll}$ \\
$t_{eclipse Ks}$ (BJD-2450000)		&	\JDOffsetONEWASPTwelveKs$^{+\JDOffsetPlusONEWASPTwelveKs}_{-\JDOffsetMinusONEWASPTwelveKs}$							& n/a																		& n/a																		& \JDOffsetTHREEWASPTwelveJointAll$^{+\JDOffsetPlusTHREEWASPTwelveJointAll}_{-\JDOffsetMinusTHREEWASPTwelveJointAll}$ 							& \JDOffsetONEWASPTwelveVariableKs$^{+\JDOffsetPlusONEWASPTwelveVariableKs}_{-\JDOffsetMinusONEWASPTwelveVariableKs}$							& \JDOffsetTHREEWASPTwelveJointVariableAll$^{+\JDOffsetPlusTHREEWASPTwelveJointVariableAll}_{-\JDOffsetMinusTHREEWASPTwelveJointVariableAll}$ \\
$t_{eclipse H}$  (BJD-2450000)		&	n/a																		& \JDOffsetONEWASPTwelveH$^{+\JDOffsetPlusONEWASPTwelveH}_{-\JDOffsetMinusONEWASPTwelveH}$							& n/a																		& \JDOffsetTWOWASPTwelveJointAll$^{+\JDOffsetPlusTWOWASPTwelveJointAll}_{-\JDOffsetMinusTWOWASPTwelveJointAll}$								& n/a																					& \JDOffsetTWOWASPTwelveJointVariableAll$^{+\JDOffsetPlusTWOWASPTwelveJointVariableAll}_{-\JDOffsetMinusTWOWASPTwelveJointVariableAll}$ \\
$t_{eclipse J}$  (BJD-2450000)		&	n/a																		& n/a																		& \JDOffsetONEWASPTwelveJ$^{+\JDOffsetPlusONEWASPTwelveJ}_{-\JDOffsetMinusONEWASPTwelveJ}$							& \JDOffsetONEWASPTwelveJointAll$^{+\JDOffsetPlusONEWASPTwelveJointAll}_{-\JDOffsetMinusONEWASPTwelveJointAll}$								& n/a																					& \JDOffsetONEWASPTwelveJointVariableAll$^{+\JDOffsetPlusONEWASPTwelveJointVariableAll}_{-\JDOffsetMinusONEWASPTwelveJointVariableAll}$ \\
$c_{1Ks}$				&	\cOneWASPTwelveKs$^{+\cOnePlusWASPTwelveKs}_{-\cOneMinusWASPTwelveKs}$										& n/a																		& n/a																		& \cOneWASPTwelveJointAll$^{+\cOnePlusWASPTwelveJointAll}_{-\cOneMinusWASPTwelveJointAll}$ 										& \cOneWASPTwelveVariableKs$^{+\cOnePlusWASPTwelveVariableKs}_{-\cOneMinusWASPTwelveVariableKs}$ 									& \cOneWASPTwelveJointVariableAll$^{+\cOnePlusWASPTwelveJointVariableAll}_{-\cOneMinusWASPTwelveJointVariableAll}$ \\
$c_{2Ks}$ ($d^{-1}$)			&	\cTwoWASPTwelveKs$^{+\cTwoPlusWASPTwelveKs}_{-\cTwoMinusWASPTwelveKs}$										& n/a																		& n/a																		& \cTwoWASPTwelveJointAll$^{+\cTwoPlusWASPTwelveJointAll}_{-\cTwoMinusWASPTwelveJointAll}$ 										& \cTwoWASPTwelveVariableKs$^{+\cTwoPlusWASPTwelveVariableKs}_{-\cTwoMinusWASPTwelveVariableKs}$									& \cTwoWASPTwelveJointVariableAll$^{+\cTwoPlusWASPTwelveJointVariableAll}_{-\cTwoMinusWASPTwelveJointVariableAll}$ \\
$c_{1H}$				&	n/a																		& \cOneWASPTwelveH$^{+\cOnePlusWASPTwelveH}_{-\cOneMinusWASPTwelveH}$ 										& n/a																		& \ParamSevenWASPTwelveJointAll$^{+\ParamSevenPlusWASPTwelveJointAll}_{-\ParamSevenMinusWASPTwelveJointAll}$								& n/a																					& \ParamSevenWASPTwelveJointVariableAll$^{+\ParamSevenPlusWASPTwelveJointVariableAll}_{-\ParamSevenMinusWASPTwelveJointVariableAll}$ \\
$c_{2H}$ ($d^{-1}$)			&	n/a																		& \cTwoWASPTwelveH$^{+\cTwoPlusWASPTwelveH}_{-\cTwoMinusWASPTwelveH}$										& n/a																		& \ParamEightWASPTwelveJointAll$^{+\ParamEightPlusWASPTwelveJointAll}_{-\ParamEightMinusWASPTwelveJointAll}$								& n/a																					& \ParamEightWASPTwelveJointVariableAll$^{+\ParamEightPlusWASPTwelveJointVariableAll}_{-\ParamEightMinusWASPTwelveJointVariableAll}$ \\
$c_{1J}$				&	n/a																		& n/a																		& \cOneWASPTwelveJ$^{+\cOnePlusWASPTwelveJ}_{-\cOneMinusWASPTwelveJ}$										& \ParamTenWASPTwelveJointAll$^{+\ParamTenPlusWASPTwelveJointAll}_{-\ParamTenMinusWASPTwelveJointAll}$									& n/a																					& \ParamTenWASPTwelveJointVariableAll$^{+\ParamTenPlusWASPTwelveJointVariableAll}_{-\ParamTenMinusWASPTwelveJointVariableAll}$ \\
$c_{2J}$ ($d^{-1}$)			&	n/a																		& n/a																		& \cTwoWASPTwelveJ$^{+\cTwoPlusWASPTwelveJ}_{-\cTwoMinusWASPTwelveJ}$										& \ParamElevenWASPTwelveJointAll$^{+\ParamElevenPlusWASPTwelveJointAll}_{-\ParamElevenMinusWASPTwelveJointAll}$								& n/a																					& \ParamElevenWASPTwelveJointVariableAll$^{+\ParamElevenPlusWASPTwelveJointVariableAll}_{-\ParamElevenMinusWASPTwelveJointVariableAll}$ \\
$\phi$	\tablenotemark{a}		&	\PhaseAbstractWASPTwelveKs$^{+\PhaseAbstractPlusWASPTwelveKs}_{-\PhaseAbstractMinusWASPTwelveKs}$						& \PhaseAbstractWASPTwelveH$^{+\PhaseAbstractPlusWASPTwelveH}_{-\PhaseAbstractMinusWASPTwelveH}$						& \PhaseAbstractWASPTwelveJ$^{+\PhaseAbstractPlusWASPTwelveJ}_{-\PhaseAbstractMinusWASPTwelveJ}$						& \PhaseAbstractWASPTwelveJointAll$^{+\PhaseAbstractPlusWASPTwelveJointAll}_{-\PhaseAbstractMinusWASPTwelveJointAll}$							& \PhaseAbstractWASPTwelveVariableKs$^{+\PhaseAbstractPlusWASPTwelveVariableKs}_{-\PhaseAbstractMinusWASPTwelveVariableKs}$						& \PhaseAbstractWASPTwelveJointVariableAll$^{+\PhaseAbstractPlusWASPTwelveJointVariableAll}_{-\PhaseAbstractMinusWASPTwelveJointVariableAll}$ \\
$\Phi_{II/I}$				&	n/a																		& n/a																		& n/a																		& n/a																					& \WidthFactorWASPTwelveVariableKs$^{+\WidthFactorPlusWASPTwelveVariableKs}_{-\WidthFactorMinusWASPTwelveVariableKs}$							& \WidthFactorWASPTwelveJointVariableAll$^{+\WidthFactorPlusWASPTwelveJointVariableAll}_{-\WidthFactorMinusWASPTwelveJointVariableAll}$\\
$\Phi_{II}$ (hours)			&	\EclipseDurationWASPTwelveKs															& \EclipseDurationWASPTwelveH															& \EclipseDurationWASPTwelveJ															& \EclipseDurationWASPTwelveJointAll																	& \EclipseDurationWASPTwelveVariableKs$^{+\EclipseDurationPlusWASPTwelveVariableKs}_{-\EclipseDurationMinusWASPTwelveVariableKs}$					& \EclipseDurationWASPTwelveJointVariableAll$^{+\EclipseDurationPlusWASPTwelveJointVariableAll}_{-\EclipseDurationMinusWASPTwelveJointVariableAll}$	\\
$T_{B Ks}$	($K$)			&	\TBrightWASPTwelveKs$^{+\TBrightPlusWASPTwelveKs}_{-\TBrightMinusWASPTwelveKs}$ 								& n/a																		& n/a																		& \TBrightWASPTwelveJointAll$^{+\TBrightPlusWASPTwelveJointAll}_{-\TBrightMinusWASPTwelveJointAll}$									& \TBrightWASPTwelveVariableKs$^{+\TBrightPlusWASPTwelveVariableKs}_{-\TBrightMinusWASPTwelveVariableKs}$								& \TBrightWASPTwelveJointVariableAll$^{+\TBrightPlusWASPTwelveJointVariableAll}_{-\TBrightMinusWASPTwelveJointVariableAll}$ \\
$T_{B H}$	($K$)			&	n/a 																		& \TBrightWASPTwelveH$^{+\TBrightPlusWASPTwelveH}_{-\TBrightMinusWASPTwelveH}$									& n/a																		& \TBrightSixWASPTwelveJointAll$^{+\TBrightSixPlusWASPTwelveJointAll}_{-\TBrightSixMinusWASPTwelveJointAll}$								& n/a																					& \TBrightSixWASPTwelveJointVariableAll$^{+\TBrightSixPlusWASPTwelveJointVariableAll}_{-\TBrightSixMinusWASPTwelveJointVariableAll}$ \\
$T_{B J}$	($K$)			&	n/a 																		& n/a																		& \TBrightWASPTwelveJ$^{+\TBrightPlusWASPTwelveJ}_{-\TBrightMinusWASPTwelveJ}$									& \TBrightNineWASPTwelveJointAll$^{+\TBrightNinePlusWASPTwelveJointAll}_{-\TBrightNineMinusWASPTwelveJointAll}$								& n/a																					& \TBrightNineWASPTwelveJointVariableAll$^{+\TBrightNinePlusWASPTwelveJointVariableAll}_{-\TBrightNineMinusWASPTwelveJointVariableAll}$ \\
$e \cos(\omega)$ \tablenotemark{a}	&	\ECosOmegaWASPTwelveKs$^{+\ECosOmegaPlusWASPTwelveKs}_{-\ECosOmegaMinusWASPTwelveKs}$								& \ECosOmegaWASPTwelveH$^{+\ECosOmegaPlusWASPTwelveH}_{-\ECosOmegaMinusWASPTwelveH}$								& \ECosOmegaWASPTwelveJ$^{+\ECosOmegaPlusWASPTwelveJ}_{-\ECosOmegaMinusWASPTwelveJ}$								& \ECosOmegaWASPTwelveJointAll$^{+\ECosOmegaPlusWASPTwelveJointAll}_{-\ECosOmegaMinusWASPTwelveJointAll}$								& \ECosOmegaWASPTwelveVariableKs$^{+\ECosOmegaPlusWASPTwelveVariableKs}_{-\ECosOmegaMinusWASPTwelveVariableKs}$								& \ECosOmegaWASPTwelveJointVariableAll$^{+\ECosOmegaPlusWASPTwelveJointVariableAll}_{-\ECosOmegaMinusWASPTwelveJointVariableAll}$\\
$e \sin(\omega)$ 			&	n/a																		& n/a																		& n/a																		& n/a																					& \ESinOmegaWASPTwelveVariableKs$^{+\ESinOmegaPlusWASPTwelveVariableKs}_{-\ESinOmegaMinusWASPTwelveVariableKs}$								& \ESinOmegaWASPTwelveJointVariableAll$^{+\ESinOmegaPlusWASPTwelveJointVariableAll}_{-\ESinOmegaMinusWASPTwelveJointVariableAll}$\\
$f_{Ks}$				& 	\fReradiationWASPTwelveKs$^{+\fReradiationPlusWASPTwelveKs}_{-\fReradiationMinusWASPTwelveKs}$							& n/a																		& n/a																		& \fReradiationWASPTwelveJointAll$^{+\fReradiationPlusWASPTwelveJointAll}_{-\fReradiationMinusWASPTwelveJointAll}$							& \fReradiationWASPTwelveVariableKs$^{+\fReradiationPlusWASPTwelveVariableKs}_{-\fReradiationMinusWASPTwelveVariableKs}$						& \fReradiationWASPTwelveJointVariableAll$^{+\fReradiationPlusWASPTwelveJointVariableAll}_{-\fReradiationMinusWASPTwelveJointVariableAll}$\\
$f_{H}$					& 	n/a																		& \fReradiationWASPTwelveH$^{+\fReradiationPlusWASPTwelveH}_{-\fReradiationMinusWASPTwelveH}$							& n/a																		& \fReradiationSixWASPTwelveJointAll$^{+\fReradiationSixPlusWASPTwelveJointAll}_{-\fReradiationSixMinusWASPTwelveJointAll}$						& n/a																					& \fReradiationSixWASPTwelveJointVariableAll$^{+\fReradiationSixPlusWASPTwelveJointVariableAll}_{-\fReradiationSixMinusWASPTwelveJointVariableAll}$\\
$f_{J}$					& 	n/a																		& n/a																		& \fReradiationWASPTwelveJ$^{+\fReradiationPlusWASPTwelveJ}_{-\fReradiationMinusWASPTwelveJ}$							& \fReradiationNineWASPTwelveJointAll$^{+\fReradiationNinePlusWASPTwelveJointAll}_{-\fReradiationNineMinusWASPTwelveJointAll}$						& n/a																					& \fReradiationNineWASPTwelveJointVariableAll$^{+\fReradiationNinePlusWASPTwelveJointVariableAll}_{-\fReradiationNineMinusWASPTwelveJointVariableAll}$\\
\enddata
\tablenotetext{a}{We account for the increased light travel-time in the system \citep{Loeb05}, and use the best-fit period for the non-precessing case reported by \citet{Campo10}.}
%\tablenotetext{b}{By assumption.}
\label{TableParams}
%\end{deluxetable}
%EMULATEAPJCHANGE
\end{deluxetable*}

\begin{figure}
\centering
\includegraphics[scale=0.35,angle=270]{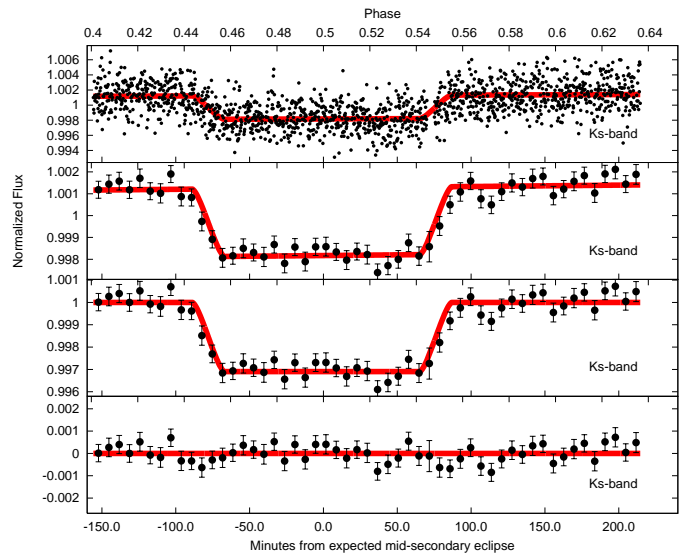}
\caption{	
		CFHT/WIRCam photometry of the secondary eclipse of WASP-12b observed in
		the Ks-band on 28 December 2009.
		The top panel shows the unbinned lightcurve with the best-fit secondary eclipse and background from our MCMC analysis of the Ks-band data with the
		fixed eclipse duration (red line).
		The second panel shows the lightcurve
		with the data binned every $\sim$7.0 minutes and again our best-fit eclipse and background.
		The third panel shows the binned data after the subtraction of
		the best-fit background, $B_f$, along with the best-fit eclipse model.
		%These middle two panels also display the
		%depth of secondary eclipse we are able to rule out at 3$\sigma$ (green dashed-line),
		%as well as the
		%depth of secondary eclipse that corresponds to zero-bond albedo and isotropic reradiation (blue line).
		The bottom panel shows the binned residuals from the best-fit model.
	}
\label{FigWASP12Ks}
\end{figure}

\begin{figure}
\centering
\includegraphics[scale=0.35,angle=270]{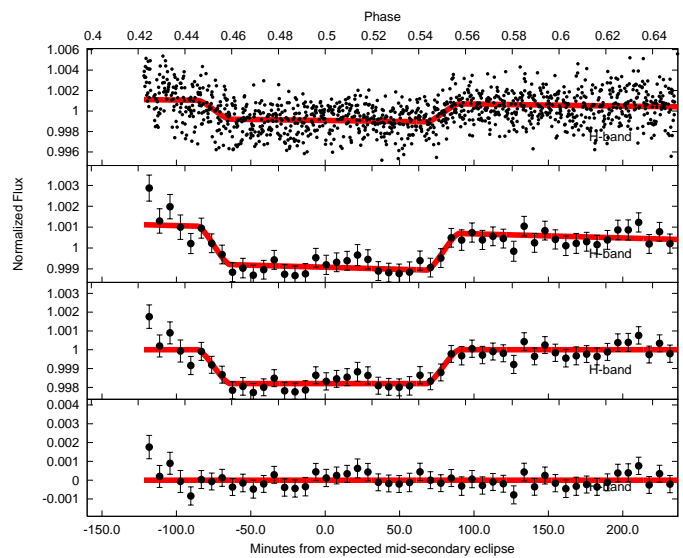}
\caption{	
		The same as figure \ref{FigWASP12Ks} except that the data is our H-band photometry obtained on 27 December 2009. 
	}
\label{FigWASP12H}
\end{figure}

\begin{figure}
\centering
\includegraphics[scale=0.35,angle=270]{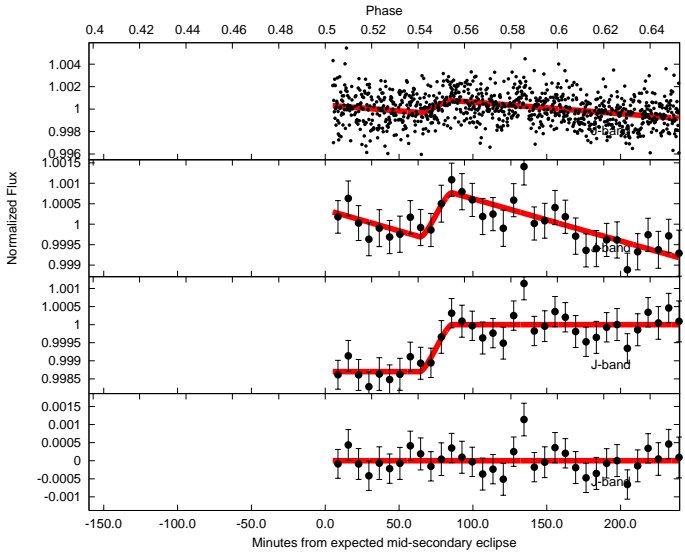}
\caption{	
		The same as figure \ref{FigWASP12Ks} except that the data is our J-band photometry obtained on 26 December 2009. Note that the photometry is a partical eclipse only,
		and starts in eclipse and extends well out of eclipse.
	}
\label{FigWASP12J}
\end{figure}

\begin{figure*}
\centering
\includegraphics[scale=0.45,angle=270]{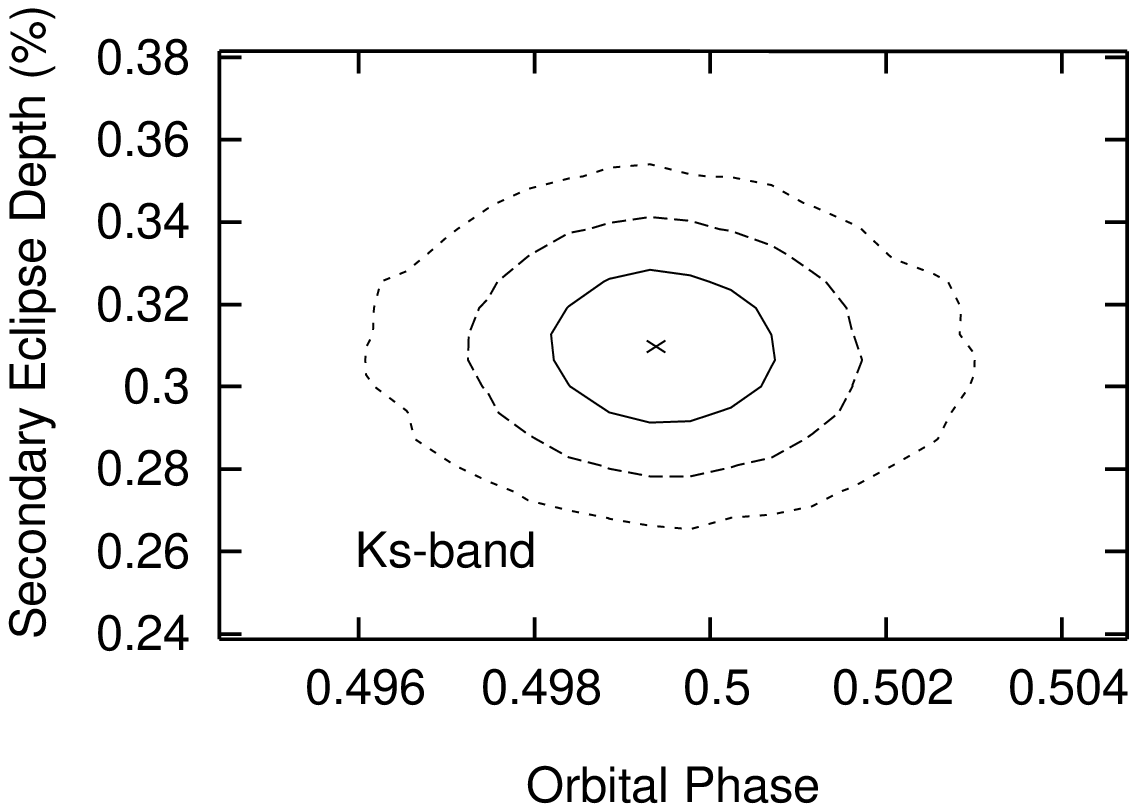}
\includegraphics[scale=0.45,angle=270]{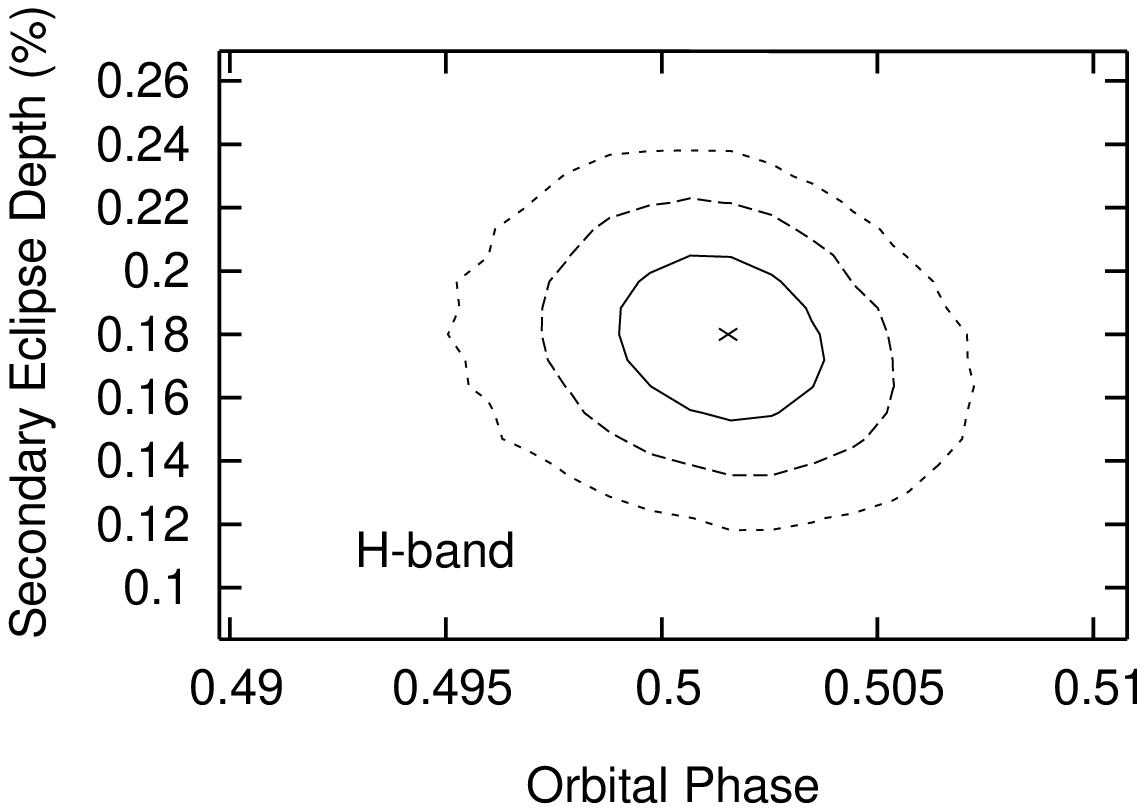}
\includegraphics[scale=0.45,angle=270]{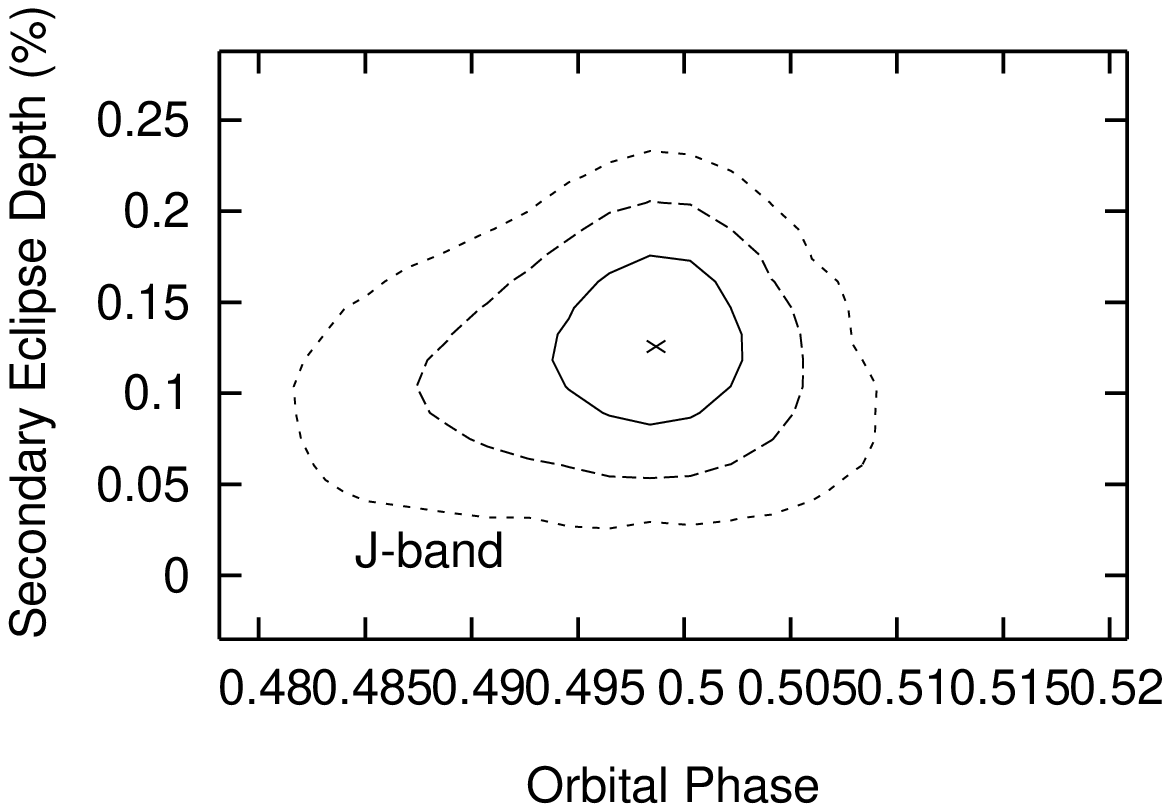}
\caption{	The 68.3\% (1$\sigma$; solid-line), 95.5\% (2$\sigma$; dashed-line), and 99.7\% (3$\sigma$; dotted-line)
		credible regions
		from our individual MCMC analyses 
		with a fixed eclipse duration of our Ks-band photometry (left), H-band photometry (middle), and J-band photometry (right).
		The ``x'' in the middle of the plots marks the best-fit point from our MCMC analyses.}
\label{FigContour}
\end{figure*}

For many of our other WIRCam data-sets we have observed
residual background trends in the reduced
data that seems to affect our target stars differently than our reference stars \citep{CrollTrESTwo,CrollTrESThree,CrollWASPThree}.
For our Ks, H \& J-band photometry these backgrounds, $B_f$,
displayed a near-linear slope.
%, while for our H-band photometry the background displayed more complicated behaviour.
We fit our Ks, H \& J-band data-sets with linear backgrounds of the form:
\begin{equation}
B_f = 1 + c_1 + c_2 dt
\end{equation}
%while for our H-band photometry we use a quadratic background:
%\begin{equation}
%B_f = 1 + c_1 + c_2 dt + c_3 dt^2
%\end{equation}
where $dt$ is the interval from the beginning of the observations
and $c_1$, $c_2$ and $c_3$ are fit parameters. 
We use Markov Chain Monte Carlo (MCMC) fitting to fit for our background as well as a secondary eclipse model calculated from the 
\citet{Mandel02}
algorithm without limb darkening.
We fit for the background, 
the depth of the secondary eclipse ($\Delta F$) and the offset
that the eclipse occurs later than the 
expected eclipse center ($t_{offset}$).
Our Markov Chain Monte Carlo method is discussed in \citet{CrollMCMC} and \citet{CrollTrESTwo}. 
We obtain our stellar and planetary parameters for WASP-12 from \citet{Hebb09}, while the planetary period
and ephemeris are obtained from \citet{Campo10} from their non-precessing best-fit.

The best-fit secondary eclipes from our individual MCMC analyses with a fixed eclipse duration 
are presented in Figures \ref{FigWASP12Ks}, \ref{FigWASP12H} and \ref{FigWASP12J}
and the best-fit eclipse parameters are presented in Table \ref{TableParams} along with associated parameters, 
such as the best-fit phase, $\phi$, and the barycentric julian date of the eclipse center in the
terrestrial time format\footnote{As calculated using the routines of \citet{Eastman10}.}, $t_{eclipse}$.
The phase dependence of these fits are presented in Figure \ref{FigContour}.
We also perform a joint analysis of the three secondary eclipses with a common offset from the eclipse
center ($t_{offset}$); the fit parameters are thus $\Delta F_Ks$, $\Delta F_H$, $\Delta F_J$, $t_{offset}$, and $c_1$, and $c_2$ 
in each band. The resulting best-fit parameters of this joint fit are listed in Table \ref{TableParams}.

We also repeat our fit for our Ks-band photometry, our highest signal-to-noise photometry, and for the joint analysis
while fitting for an additional parameter - the duration of the secondary eclipse, $\Phi_{II}$.
We parameterize this by the duration of the eclipse divided by the duration of the transit, $\Phi_{II/I}$,
using the duration of the transit ($\Phi_I$$\sim$2.93 $h$) reported by \citet{Hebb09}. The results from this fit are presented in 
Table \ref{TableParams}. We do not fit our J-band or H-band data individually with this additional parameter, $\Phi_{II}$,
as the J-band data is a partial eclipse and thus the duration of the secondary eclipse is degenerate with an offset of the eclipse center,
and the H-band data suffers from additional time-correlated systematics that could lead to erroneous conclusions.

\section{Discussion}

 We strongly detect all three secondary eclipses in the three near-infrared bands that we observed in.
The individual analyses of our three eclipses confirm
that all three secondary eclipses are fit with a consistent phase (Table \ref{TableParams}); thus the 
best-fit parameters from our joint analysis are similar to the parameters returned by the analyses
of the individual eclipses. We therefore quote the results of the joint
analysis below. The best-fit eclipse depths from our joint analysis is  
\FpOverFStarPercentAbstractWASPTwelveJointAll$^{+\FpOverFStarPercentAbstractPlusWASPTwelveJointAll}_{-\FpOverFStarPercentAbstractMinusWASPTwelveJointAll}$\% in Ks-band,
\ParamSixWASPTwelveJointAll$^{+\ParamSixPlusWASPTwelveJointAll}_{-\ParamSixMinusWASPTwelveJointAll}$\% in H-band and
\ParamNineWASPTwelveJointAll$^{+\ParamNinePlusWASPTwelveJointAll}_{-\ParamNineMinusWASPTwelveJointAll}$\% in J-band.

\subsection{Eccentricity and Precession of WASP-12b}
\label{SecEccentricity}

%EMULATEAPJCHANGE
\begin{deluxetable*}{cccc}
%\begin{deluxetable}{cccc}
\tabletypesize{\footnotesize}
\tablecaption{WASP-12b orbital parameters}
\tablehead{
\colhead{Parameter} 			& \colhead{Precessing}	& \colhead{Non-Precessing} 				& \colhead{Non-Precessing Case without }\\	
\colhead{} 			& \colhead{Case}	& \colhead{Case} 				& \colhead{the \citet{LopezMorales10} eclipse }\\	
}
\startdata
$P$ (days)				& \PrecessionOneLikely$^{+\PrecessionOnePlus}_{-\PrecessionOneMinus}$				& \PrecessionOneNoPrecessLikely$^{+\PrecessionOneNoPrecessPlus}_{-\PrecessionOneNoPrecessMinus}$				& \PrecessionOneNoPrecessNoLopezLikely$^{+\PrecessionOneNoPrecessNoLopezPlus}_{-\PrecessionOneNoPrecessNoLopezMinus}$ 			\\
$e$					& \PrecessionTwoLikely$^{+\PrecessionTwoPlus}_{-\PrecessionTwoMinus}$				& \PrecessionTwoNoPrecessLikely$^{+\PrecessionTwoNoPrecessPlus}_{-\PrecessionTwoNoPrecessMinus}$				& \PrecessionTwoNoPrecessNoLopezLikely$^{+\PrecessionTwoNoPrecessNoLopezPlus}_{-\PrecessionTwoNoPrecessNoLopezMinus}$ 			\\
$T_o$ (BJD-2450000)			& \PrecessionZeroLikely$^{+\PrecessionZeroPlus}_{-\PrecessionZeroMinus}$			& \PrecessionZeroNoPrecessLikely$^{+\PrecessionZeroNoPrecessPlus}_{-\PrecessionZeroNoPrecessMinus}$				& \PrecessionZeroNoPrecessNoLopezLikely$^{+\PrecessionZeroNoPrecessNoLopezPlus}_{-\PrecessionZeroNoPrecessNoLopezMinus}$ 			\\
$\omega_o$ ($^{o}$)			& \PrecessionThreeLikely$^{+\PrecessionThreePlus}_{-\PrecessionThreeMinus}$ \tablenotemark{a}	& \PrecessionThreeNoPrecessLikely$^{+\PrecessionThreeNoPrecessPlus}_{-\PrecessionThreeNoPrecessMinus}$ \tablenotemark{a}	& \PrecessionThreeNoPrecessNoLopezLikely$^{+\PrecessionThreeNoPrecessNoLopezPlus}_{-\PrecessionThreeNoPrecessNoLopezMinus}$ \tablenotemark{a}	\\
$\dot{\omega}$ ($^{o}$ d$^{-1}$)	& \PrecessionFourLikely$^{+\PrecessionFourPlus}_{-\PrecessionFourMinus}$			& 0.0 \tablenotemark{b}														& 0.0 \tablenotemark{b} 												\\
$e$cos$\omega_o$			& \PrecessionECosOmegaLikely$^{+\PrecessionECosOmegaPlus}_{-\PrecessionECosOmegaMinus}$ 	& \PrecessionECosOmegaNoPrecessLikely$^{+\PrecessionECosOmegaNoPrecessPlus}_{-\PrecessionECosOmegaNoPrecessMinus}$		& \PrecessionECosOmegaNoPrecessNoLopezLikely$^{+\PrecessionECosOmegaNoPrecessNoLopezPlus}_{-\PrecessionECosOmegaNoPrecessNoLopezMinus}$ 	\\
$e$sin$\omega_o$			& \PrecessionESinOmegaLikely$^{+\PrecessionESinOmegaPlus}_{-\PrecessionESinOmegaMinus}$ 	& \PrecessionESinOmegaNoPrecessLikely$^{+\PrecessionESinOmegaNoPrecessPlus}_{-\PrecessionESinOmegaNoPrecessMinus}$		& \PrecessionESinOmegaNoPrecessNoLopezLikely$^{+\PrecessionESinOmegaNoPrecessNoLopezPlus}_{-\PrecessionESinOmegaNoPrecessNoLopezMinus}$ 	\\
$\chi^{2}$				& \PrecessionChiSquaredLikely									& \PrecessionChiSquaredNoPrecessLikely												& \PrecessionChiSquaredNoPrecessNoLopezLikely 											\\
BIC					& \PrecessionBICLikely										& \PrecessionBICNoPrecessLikely													& \PrecessionBICNoPrecessNoLopezLikely 											\\
\enddata
\tablenotetext{a}{These distributions are bimodal with strong peaks at $\omega$$\sim$90$^{o}$ and -90$^{o}$ (where $\cos$$\omega$$\sim$0.)}
\tablenotetext{b}{By definition.)}
\label{TablePrecession}
%\end{deluxetable}
%EMULATEAPJCHANGE
\end{deluxetable*}

 The best-fit phase of the joint analysis is $\phi$=\PhaseAbstractWASPTwelveJointAll$^{+\PhaseAbstractPlusWASPTwelveJointAll}_{-\PhaseAbstractMinusWASPTwelveJointAll}$.
The resulting limit on the eccentricity, $e$, and argument of periastron, $\omega$, is 
$e$cos$\omega$=\ECosOmegaWASPTwelveJointAll$^{+\ECosOmegaPlusWASPTwelveJointAll}_{-\ECosOmegaMinusWASPTwelveJointAll}$, a result
that is consistent with a circular orbit and the \citet{Campo10} results.
This value is inconsistent, however, with the \citet{LopezMorales10} $e$cos$\omega$ result. The 
discrepancy between the 
\citet{LopezMorales10} result and that of 
\citet{Campo10} and our own
could be due to WASP-12b precessing - we explore this possibility below.

\begin{figure}
\centering
\includegraphics[scale=0.45,angle=270]{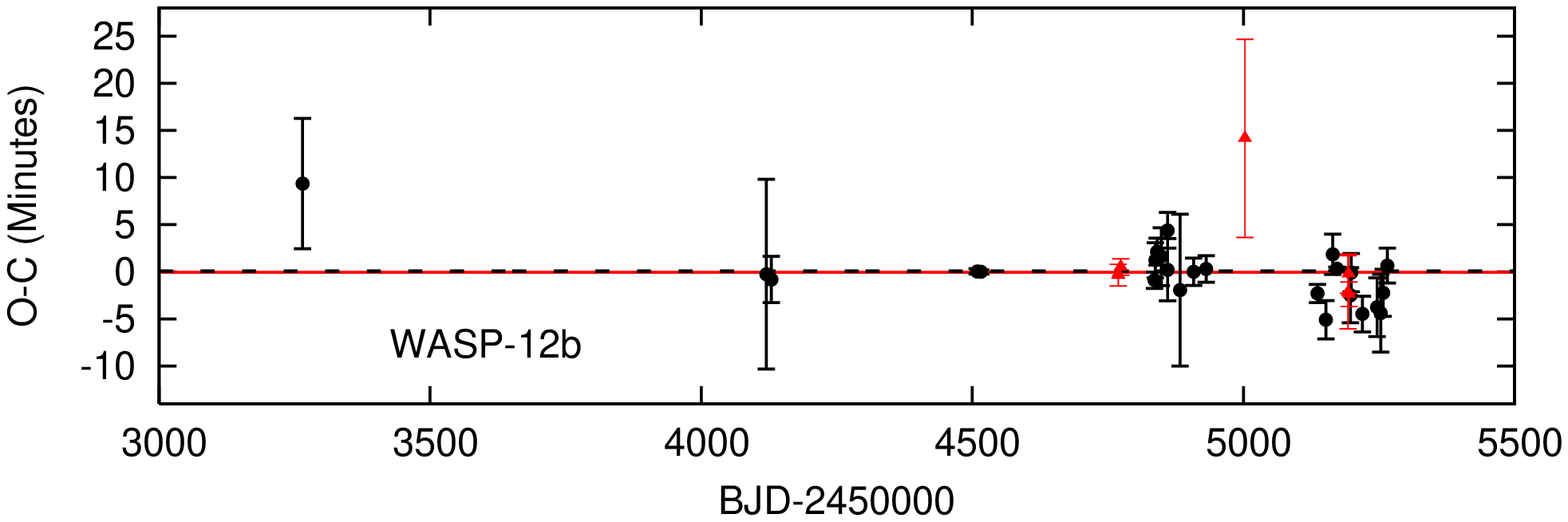}
\includegraphics[scale=0.45,angle=270]{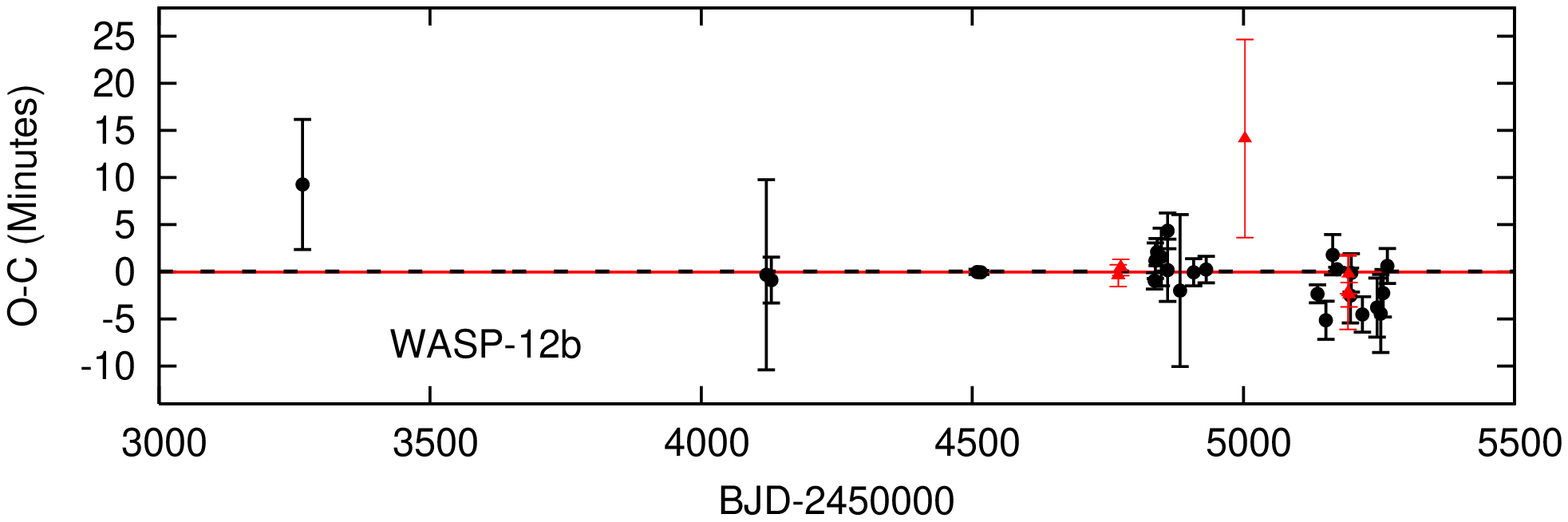}
\caption{	Transit (black points) and eclipse (red points) times for WASP-12b
		compared to the best-fit orbital models for the 
		precessing case (top),
%		the non-precessing case (middle), and the non-precessing case with the \citet{LopezMorales10} point excluded.
		and the non-precessing case (bottom).
		The best-fit models for the transit times (dotted black line)
		and the eclipse times (solid red line) are also shown.
		Both diagrams show the observed-minus-calculated (O-C) times from
		a linear ephemeris calculated using $T_o$ and $P$; the secondary
		eclipse O-C times are compared to a calculated eclipse centre of $T_o$ + $\frac{P}{2}$.
		The eclipse points in the two panels are from left-to-right, the \citet{Campo10} Spitzer/IRAC eclipses (BJD-2450000$\sim$4750),
		the \citet{LopezMorales10} eclipse (BJD-2450000$\sim$5000), while the last
		three red-points are the eclipse times reported here (BJD-2450000$\sim$5200).
}
\label{FigPrecession}
\end{figure}

% MORE ON THE ECCENTRICITY HERE, using the argument of periastron.
% Tidal Q, ``puffed-up'' radius. Is it consistent with the radial velocity limit.
% If this eccentricity is real, than we can also expect transit and eclipse
% timing variations for WASP-12 on the order of $\frac{eP}{\pi}$ \citep{RagozzineWolf09}, or 2.5 minutes.

% Using this inferred eccentricity, we are also able to estimate the ratio of the tidal luminosity to the stellar insolation, $\frac{\dot{E_{tide}}}{\dot{E_{insolation}}}$,
% following the arguments of \citet{Liu08}. 
% Assuming a canonical tidal quality paramter of, $Q$=10$^{6}$, we estimate:
% $\frac{\dot{E_{tide}}}{\dot{E_{insolation}}}$$\sim$7\%. This esimate is highly dependent on, $Q$, which is very uncertain
% for this planet as well as other hot Jupiters.

	\citet{Campo10} performed an analysis of the reported transit times and secondary eclipse times and presented tentative evidence
that WASP-12b may be precessing at an observable rate, $\dot{\omega}$ = 0.02 $\pm$ 0.01$^{o}$ d$^{-1}$, with a period as short as 40 years.
The primary evidence for the precession was the 
ground-based secondary eclipse detection of \citet{LopezMorales10}, which
occured late by approximately $\sim$15 minutes (at a phase of 
$\phi$=0.5100$^{+0.0072}_{-0.0061}$ using the \citealt{Hebb09}
ephemeris and period).

We repeat the \citet{Campo10} precession
analysis adding in our three secondary eclipse detections. We summarize the \citet{Campo10} precession
model that we employ here. The mid-transit time of the $N^{th}$ transit, $T_N$, 
in our precessing model
is predicted to occur at:
\begin{equation}
T_N = T_o + P_s N - \frac{e P_a}{\pi} (cos \omega_N - cos \omega_o).
\label{EqunPrecession}
\end{equation} 
$T_o$ and $\omega_o$ are the transit time and argument of periastron
at the reference epoch, $\omega_N$ is the argument of periastron of the $N^{th}$ transit,
$P_s$ is the sidereal period, $P_a$ is the period between successive periastron passages, and $e$ has 
already been defined
as the eccentricity.
$P_a$ is not an independent variable, but is related to the sidereal
period, $P_s$, and the constant precession rate, $\dot{\omega}$:
$P_a = \frac{P_s}{1-P_s\frac{\dot{\omega}}{2\pi}}$.
The argument of periastron of the $N^{th}$ transit is simply $\omega_N$ = $\dot{\omega} (T_N - T_o) + \omega_o$.
Equation \ref{EqunPrecession} is solved iteratively for $T_N$ after it is 
expanded to fifth order in $e$ 
(as shown in equation (22) of \citealt{RagozzineWolf09}). 
We fit the radial velocity data from \citet{Hebb09} 
and \citet{Husnoo10}\footnote{\citet{Husnoo10} argue that there may be correlated
red noise in the \citet{Hebb09} radial-velocity data, possibly due to a systematic offset
in the RV zero-point from night to night. As a result we scale-up
the errors for the \citet{Hebb09} data by a factor of 8 and those of \citet{Husnoo10} by a factor of 2 to account for possible
offsets between these two data-sets. We refer the reader to \citet{Husnoo10} for further discussion.},
and the transits listed in Table 2 of \citet{Campo10} as well as four additional,
recent transits\footnote{The additional transits have mid-transit times (HJD) of 2455246.77604$\pm$0.00217 (A. Gibson, TRESCA), 2455253.32414$\pm$0.00287 (F. Lomoz, TRESCA), 2455257.69131 (G. Haagen, TRESCA), and 2455265.33327$\pm$0.00129 (H. Kucakova, TRESCA).}
from the Exoplanet Transit database \citep{Poddany10}
and our own secondary eclipse data along with those of
\citet{LopezMorales10}, and \citet{Campo10}. We exclude the in-transit radial
velocity data as we do not model for the Rossiter-McLaughlin effect \citep{GaudiWinn07}.
We follow \citet{Campo10}, and quote the \citet{LopezMorales10} eclipse point 
that results from the combined photometry from 1.5 eclipses, 
at a single epoch
halfway between their observations (HJD$\sim$2455002.8560 $\pm$ 0.0073).
$T_N$ of course gives the transit
time to compare to the data, we use $e$, $\omega_N$, $T_N$ and $P_a$ to calculate the eclipse times, and $\omega(t)$ to calculate
the radial velocity values. We use the MCMC techniques explained above to calculate the best-fit precessing
model, and non-precessing models, except that we fit for $e$$\cos$$\omega$ and $e$$\sin$$\omega$, instead of $e$ and $\omega$,
as
$\omega$ is poorly constrained as the eccentricity approaches zero.

%\begin{figure}
%\centering
%\includegraphics[scale=0.45,angle=270]{HIST62_HISTO.eps}
%\caption{	Marginalized likelihood for the eccentricity of WASP-12b from the non-precessing MCMC chain with
%		the \citet{LopezMorales10} point excluded.
%		The best-fit value is given with the solid vertical line (and is nearly indistinguishable from zero),
%		while the 68\% credible region 
%		is indicated by the dotted vertical line.
%}
%\label{FigHisto}
%\end{figure}

\begin{figure*}
\centering
\includegraphics[scale=0.60,angle=270]{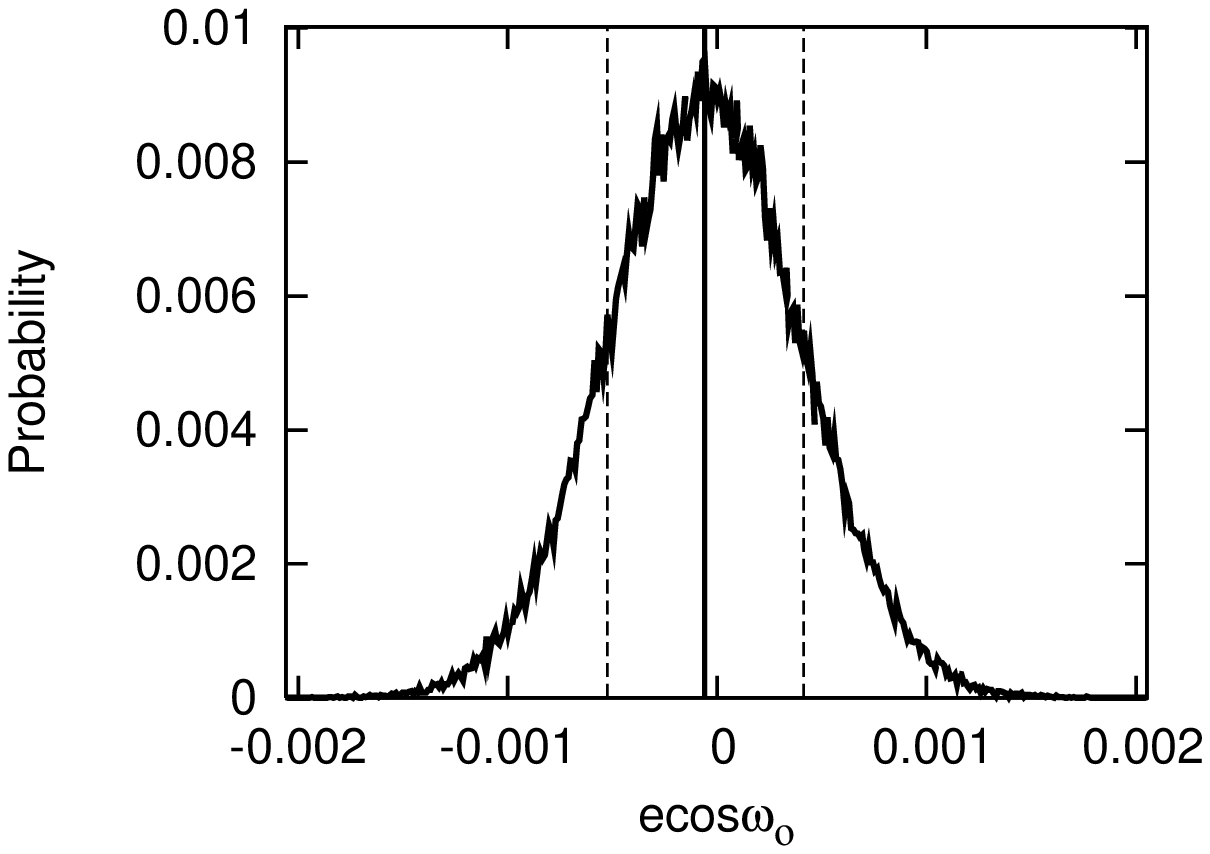}
\includegraphics[scale=0.60,angle=270]{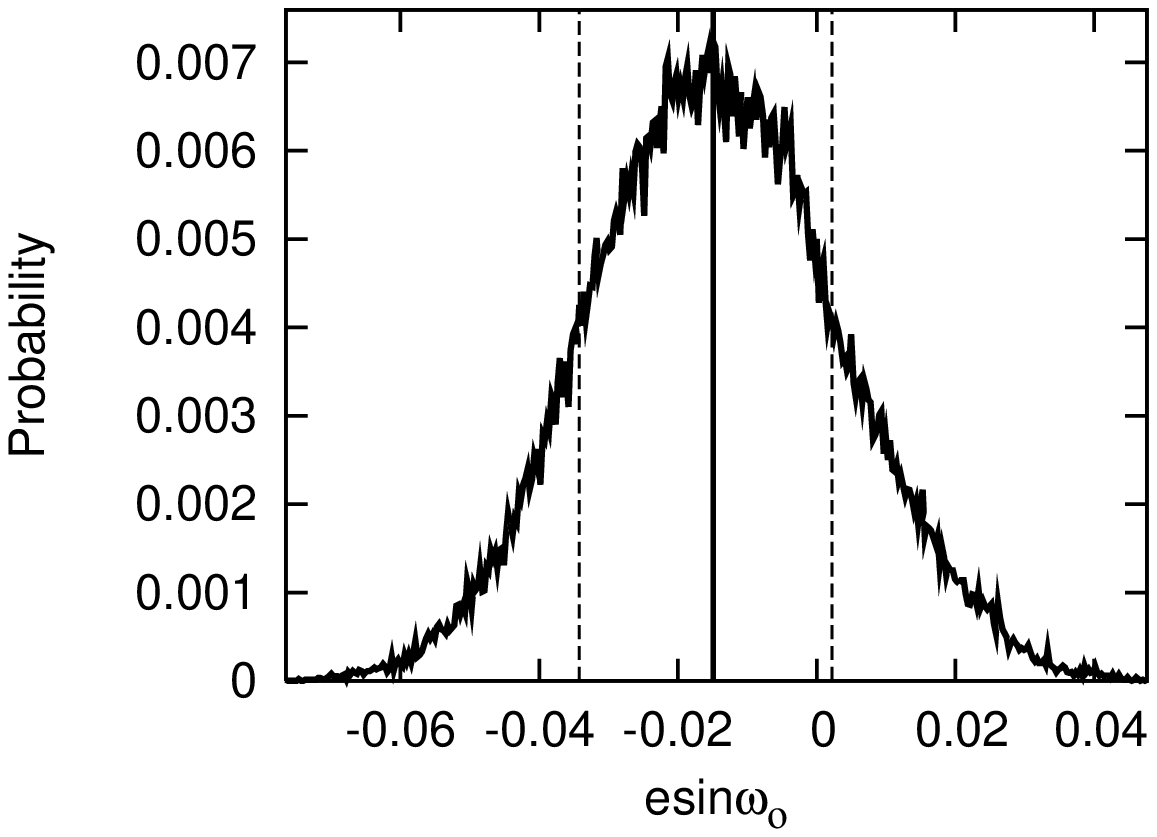}
\includegraphics[scale=0.60,angle=270]{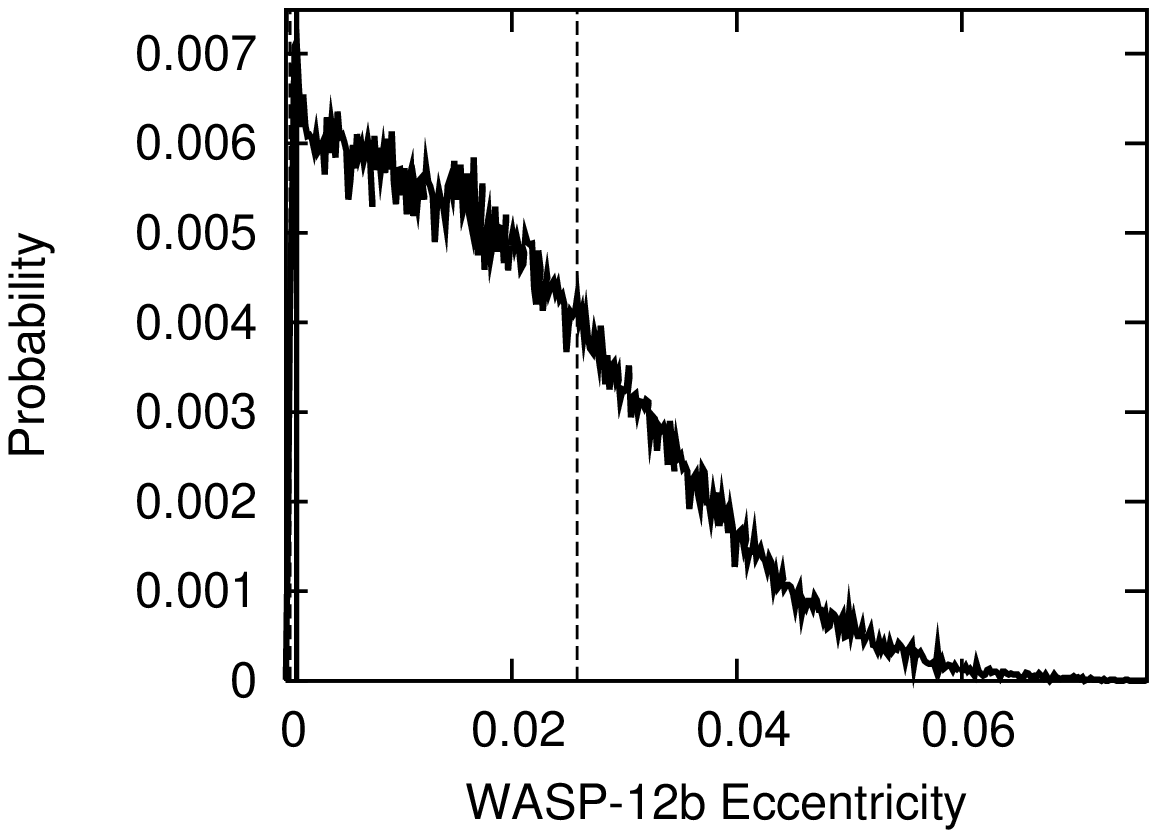}
\includegraphics[scale=0.64,angle=270]{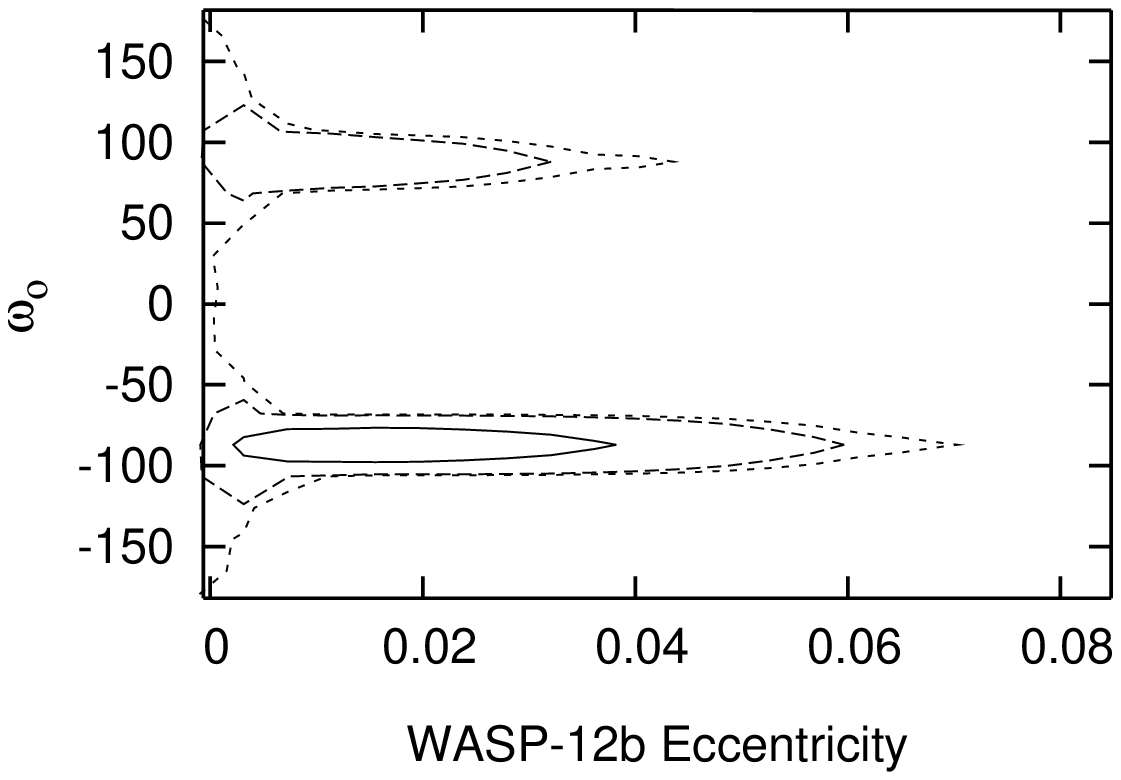}
\caption{	Top-left, top-right and bottom-left panels: 
		Marginalized likelihood for WASP-12b's $e$$\cos$$\omega$, $e$$\sin$$\omega$
		and its eccentricity from the non-precessing MCMC chain with
		the \citet{LopezMorales10} point excluded.
		The best-fit value for each panel is given with the solid vertical line (for the 
		bottom-left panel this value is nearly indistinguishable from zero),
		while the 68\% credible region 
		is indicated by the dotted vertical line.
		Bottom-right panel: Contour parameter showing the eccentricity, $e$,
		and the argument or periastron, $\omega$, of WASP-12b again from the
		same MCMC chain. 
		The 68.3\% (1$\sigma$; solid-line), 95.5\% (2$\sigma$; dashed-line), and 99.7\% (3$\sigma$; dotted-line)
		credible regions are indicated.
}
\label{FigHisto}
\end{figure*}

 We plot our precessing and non-precessing best-fit models in
Figure \ref{FigPrecession} and present the MCMC results in Table \ref{TablePrecession}.
The best-fit models with and without precession are similar. The best-fit precessing model features
a very small rate of precession ($\dot{\omega}$ = \PrecessionFourLikely$^{+\PrecessionFourPlus}_{-\PrecessionFourMinus}$$^{o}$ d$^{-1}$),
that barely provides a superior fit once the extra degrees of freedom are taken into account (a Bayesian Information Criterion\footnote{For the Bayesian Information Criterion \citep{Liddle07} lower-values indicate superior fits corrected for the number of free parameters: $BIC$=$\chi^2$ + $k$ln$N$, where $k$ is the number of free parameters and $N$ is the number of data points.}
of $BIC$=\PrecessionBICLikely \ for the precessing case, compared
to $BIC$=\PrecessionBICNoPrecessLikely \ for the non-precessing case).  
Thus there is not convincing evidence at this date that WASP-12b is precessing.

%It also appears likely that the systematic errors associated with the timing (and thus possibly the depth) 
%of the \citet{LopezMorales10} eclipse detection have been underestimated. 
%We also refit the non-precessing case with this eclipse excluded, 
%proofdif
Given that the timing offset of the \citet{LopezMorales10} eclipse detection may be suspect,
we also refit the non-precessing case with this eclipse excluded, 
% and plot the best-fit model in the bottom panel of Figure \ref{FigPrecession},
and present the MCMC results in Table \ref{TablePrecession}.
The distribution of eccentricity values from our MCMC chain without the \citet{LopezMorales10} eclipse
is non-gaussian (the bottom left panel of Figure \ref{FigHisto}) and favours a near-zero eccentricity
with a tail to higher eccentricity values;
this limit is 
$e$=\PrecessionTwoNoPrecessNoLopezLikely$^{+\PrecessionTwoNoPrecessNoLopezPlus}_{-\PrecessionTwoNoPrecessNoLopezMinus}$.
This is due to the fact that although 
the $e$$\cos$$\omega_o$ values for WASP-12b are well-constrained from the radial-velocity data
and the combination of the timing of the eclipses and transits (the top-left panel of Figure \ref{FigHisto}),
the $e$$\sin$$\omega_o$ values are not well-constrained
and thus higher eccentricity values are allowed (the top-right panel of Figure \ref{FigHisto})
for an argument of periastron where
$\cos$$\omega_o$$\sim$0 at $\omega_o$$\sim$90$^{o}$ and -90$^{o}$ (as can be seen in the 
contour plot in the bottom-right panel of Figure \ref{FigHisto}). 
Although we are not able to rule out
higher eccentricity values for WASP-12b with high confidence,
the orbit of WASP-12b is likely
circular; thus WASP-12b is no longer an outlier from the expectation of the timescale
of tidal circularization for close-in giant exoplanets.
The above analysis would be improved by including an a
priori constraint on $e$$\sin$$\omega_o$ using the eclipse duration values from 
our own eclipses and the \citet{Campo10} Spitzer/IRAC eclipses. Unfortunately, although \citet{Campo10} indicate that 
their best-fit eclipse durations are similar to that of the transits and should thus place a tight
constrain on $e$$\sin$$\omega_o$ near
zero, \citet{Campo10} do not formally fit for the duration of the eclipse 
and do not include the associated uncertainties. We discuss the implications of fitting our own
eclipse durations below.

\subsection{A longer duration secondary eclipse; possible signs of material stripped from the planet?}
\label{SecDisk}

\begin{figure}
\centering
\includegraphics[scale=0.35,angle=270]{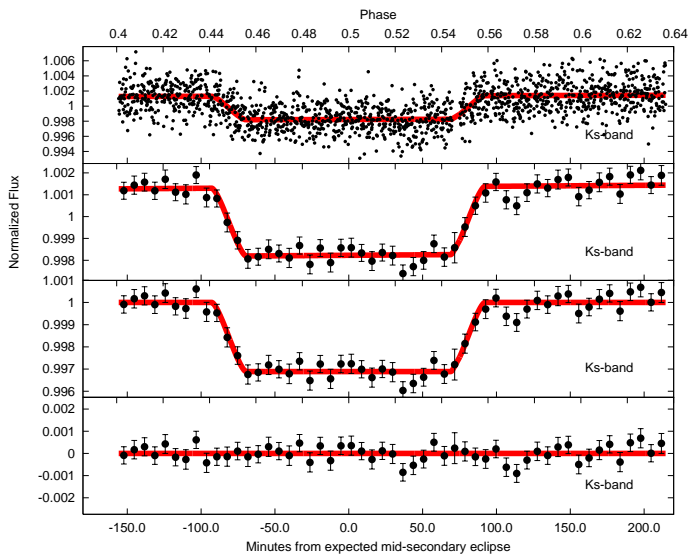}
\caption{	
		The same as figure \ref{FigWASP12Ks} except the best-fit model is our variable eclipse duration model for our Ks-band photometry.
	}
\label{FigWASP12KsVariable}
\end{figure}

 We also fit our Ks-band photometry and our joint J, H \& Ks-band photometry 
with an eclipse model with the eclipse duration as a free parameter.
Our best-fit Ks-band variable eclipse duration fit is presented in Figure \ref{FigWASP12KsVariable}. Our variable eclipse duration fit
does argue for a marginally wider secondary eclipse than transit:
$\Phi_{II/I}$ = \WidthFactorWASPTwelveVariableKs$^{+\WidthFactorPlusWASPTwelveVariableKs}_{-\WidthFactorMinusWASPTwelveVariableKs}$,
although this result is only significant at the \WidthFactorSigmaWASPTwelveVariableKs$\sigma$-level.
The associated eclipse duration is 
$\Phi_{II}$ = \EclipseDurationWASPTwelveVariableKs$^{+\EclipseDurationPlusWASPTwelveVariableKs}_{-\EclipseDurationMinusWASPTwelveVariableKs}$ hours,
longer than the $\sim$2.93 hour optical transit found by \citet{Hebb09}, and longer than 
the similar duration IRAC eclipses found by \citet{Campo10}.
That the data suggests a wider secondary eclipse than our best-fit model can be seen
in the ingress and egress of our Ks-band 
photometry (Figure \ref{FigWASP12Ks}). 

Our joint analysis of our J, H \& Ks-band data also argues for a marginally wider secondary eclipse
than transit: 
$\Phi_{II/I}$ = \WidthFactorWASPTwelveJointVariableAll$^{+\WidthFactorPlusWASPTwelveJointVariableAll}_{-\WidthFactorMinusWASPTwelveJointVariableAll}$, or that the duration of
the eclipse is $\Phi_{II}$ = \EclipseDurationWASPTwelveJointVariableAll$^{+\EclipseDurationPlusWASPTwelveJointVariableAll}_{-\EclipseDurationMinusWASPTwelveJointVariableAll}$ hours.
As our J-band data is a partial eclipse, it has no ability to constrain the eclipse duration on its own. 
Similarly, as our H-band data suffers from significant systematics prior to and during ingress, its ability to constrain
the eclipse duration is compromised; in fact, the systematics at the beginning of the H-band photometry
that manifest themselves as a sharp decrease in flux, can be
well-fit by a significantly wider, and deeper secondary eclipse that is unlikely to be physical.
These facts, combined with a visual inspection of 
Figures \ref{FigWASP12Ks}-\ref{FigWASP12J},
suggests that the wider secondary eclipse for our joint analysis, is in-fact dominated by our Ks-band photometry and 
the longer duration eclipse may not be
credible for the joint analysis.  

 Our Ks-band photometry is best-fit with
a wider secondary eclipse than expected:
$\Phi_{II/I}$=\WidthFactorWASPTwelveVariableKs$^{+\WidthFactorPlusWASPTwelveVariableKs}_{-\WidthFactorMinusWASPTwelveVariableKs}$.
The first possibility to explain this wider than expected
eclipse is systematic time-correlated, red-noise in our photometry, which would not be unexpected
as the eclipse is only wider than expected at less than the 3$\sigma$ level.
Another possibility for this wider eclipse is that the planet has
a small eccentricity ($e$$\sin$$\omega$=\ESinOmegaWASPTwelveVariableKs$^{+\ESinOmegaPlusWASPTwelveVariableKs}_{-\ESinOmegaMinusWASPTwelveVariableKs}$). 
We have already presented strong evidence that the eccentricity of WASP-12b is quite likely 
near zero in $\S$\ref{SecEccentricity} and the \citet{Campo10} Spitzer/IRAC eclipse photometry
does not feature a longer duration secondary eclipse. Also, although the $e$sin$\omega$ of the planet is less well constrained
in $\S$\ref{SecEccentricity}, an $e$sin$\omega$ value necessary to explain our longer duration eclipse
can be ruled out at several sigma and thus we 
find this possibility uncompelling.

Another possibility -- perhaps the most intriguing possibility --
is that if this apparently wider secondary eclipse
is not due to
systematic effects or due to a small $e$$\sin$$\omega$
for WASP-12b, then it could be due to radiation from gas
that is escaping from the planet and possibly
forming a cicumstellar disk. 
The latter was
predicted by \citet{Li10}, while the former was arguably recently confirmed by \citet{Fossati10} through observations that
WASP-12b displayed increased transit depths in the UV with COS/HST.
An eclipse of
this duration could argue for material surrounding the planet
with a projected radius
that is approximately \RWideWASPTwelve \ times
the optical radius of the planet, or at a radius of \RWideJupiter \ $R_{Jupiter}$,
and would thus argue for material emitting radiation that is exceeding the Roche lobe, and streaming from
the planet. 
This emission could be due to CO $\sim$2.292 bandhead emission, as predicted by \citet{Li10}, although the material
around the planet should be cooler than the $\sim$4000-5000 K temperatures they predicted for the cirumstellar disk and will
thus result in reduced emission.
In the ``accretion stream'' hypothesis, advocated by \citep{Lai10},
the material streaming from the planet towards the star may be highly localized in a line passing through the inner
Lagrangian point. The extra emission from this stream would be obscured
by the star earlier than the planet during eclipse
ingress and later than the planet during eclipse egress.
Such a scenario is arguably favoured over simply a 
sphere of evaporating material \RWideJupiter \ $R_{Jupiter}$ in radius, as in the accretion stream scenario
the emission will arise from a smaller surface area; otherwise 
the Ks-band brightness temperature of the planet
would have to be anomalously low, given that the 
$\Delta F_{Ks}$=\FpOverFStarPercentAbstractWASPTwelveJointAll$^{+\FpOverFStarPercentAbstractPlusWASPTwelveJointAll}_{-\FpOverFStarPercentAbstractMinusWASPTwelveJointAll}$\%
that we observe would have to be a combination of the planet and the enveloping
material that would have a much larger surface area of emitting material
than the planet itself.
% -- such a sphere of evaporating material  arguably disfavoured 
% require an anamously low Ks-band brightness temperature
%for the planet.

Alternatively,
this wider eclipse could be interpreted as the planet passing behind a circumstellar disk that is optically thick
at these wavelengths and extends marginally from the star (at least $\sim$1.11 times % $\sim$\WidthFactorWASPTwelveVariableKs \ times
the stellar radius), and therefore
obscures the planet earlier and later than expected. If the disk is optically thick it will have to be due to gas
opacity, as the temperature of the disk will be well above the dust sublimation temperature, and 
the temperature and 
density of the disk will have to be high enough for the material to be largely photoionized
to avoid the ``opacity gap'' \citep{Thompson05}. 
The disk would also
have to be optically thick at around 2 $\mu m$, but not at the longer wavelengths probed by Spitzer/IRAC 
(3.6 to 8.0 $\mu m$) as the durations of the eclipses are not discrepant from
the expected duration in the \citet{Campo10} photometry. The ``accretion stream'' hypothesis is arguably less contrived, but
the observed eclipse duration could also result from a combination of both scenarios.

An obvious way to differentiate between these two scenarios would be observations
of WASP-12 during transit in the Ks-band band. 
If WASP-12 is surrounded by a
circumstellar disk that is emitting in Ks-band then the transit duration will increase.
If there is material surrounding WASP-12b then its transit will be of the 
expected duration if the material is optically thin, and the transit will display an increased
depth if the material is optically thick. 
We plan to perform such follow-up observations of the transit and eclipse of 
WASP-12b in the Ks and H-bands to differentiate between these various scenarios,
and to confirm the near-zero eccentricity of WASP-12b. Until such follow-up observations take place we emphasize that
our Ks-band photometry is best-fit with a wider eclipse at less than the 3$\sigma$ level.

% Others include that temperature variations on the exoplanet could manifest themselves during ingress and egress;
% this seems unlikely as it would require the hottest part of the planet to be near the limb of the planet.
% The most intriguing possibility is that these variations in ingress and egress could be due to the fact that the
% planet is prolate as has been suggested
% by \citet{Li10}, and has
% filled a large fraction of its Roche lobe. This possibility will be explored in an upcoming publication.

%Thus does appear to be an offset from the period of WASP-12b as returned by secondary eclipse and transit times, and the radial velocity.

\subsection{The properties of WASP-12b's atmosphere}

\begin{figure}
\centering
\includegraphics[scale=0.65,angle=270]{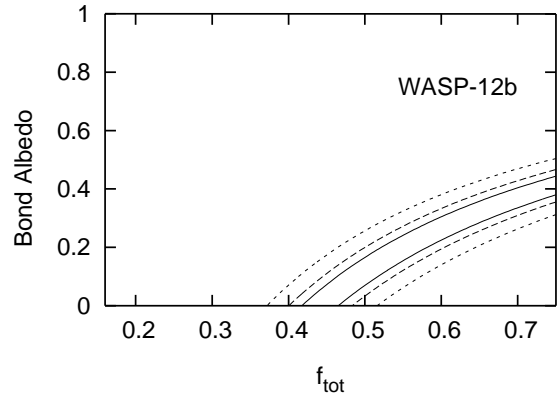}
\caption{	
		1$\sigma$ (solid-lines), 2$\sigma$ (dashed-lines), and 3$\sigma$
		(dotted-lines) constraints on the Bond albedo and reradiation factor, $f_{tot}$
		from our Ks, H \& J-band secondary eclipse observations of WASP-12b.
		% $f$ = 0.25 corresponds to isotropic reradiation, and thus $f$ values
		% less than 0.25 are unphysical unless there are absorption bands at these wavelengths.
	}
\label{FigBondReradiation}
\end{figure}

	Our measurements of the thermal emission of WASP-12b allow us to constrain the characteristics of its atmosphere,
including:
its Bond albedo, the level of redistribution of heat from the day to the nightside at various depths,
and the planet's dayside bolometric luminosity.
We parameterize the level of redistribution by the reradiation factor,
$f$, following the \citet{LopezMoralesSeager07} definition (i.e. 
$f$=$\frac{1}{4}$ denotes isotropic reradiation, while $f$=$\frac{1}{2}$
denotes redistribution and reradiation from the dayside only).
Our eclipse depths are consistent with a range of Bond albedos, $A_B$,
and overall day to nightside redistribution of heat, $f_{tot}$ (Figure \ref{FigBondReradiation}).
If we assume a Bond albedo near zero, consistent with observations of other hot Jupiters \citep{Charbonneau99,Rowe08}
and with model predictions \citep{Burrows08}, the best-fit reradiation factor, $f_{tot}$, that results from our three near-infrared
eclipse measurements is
$f_{tot}$ = \fReradiationWASPTwelveALL$^{+\fReradiationPlusWASPTwelveALL}_{-\fReradiationMinusWASPTwelveALL}$.
This suggests that the dayside of WASP-12b reradiates most of the incident stellar flux
without redistributing it to the nightside. 

\begin{figure}
\centering
\includegraphics[scale=0.49,angle=270]{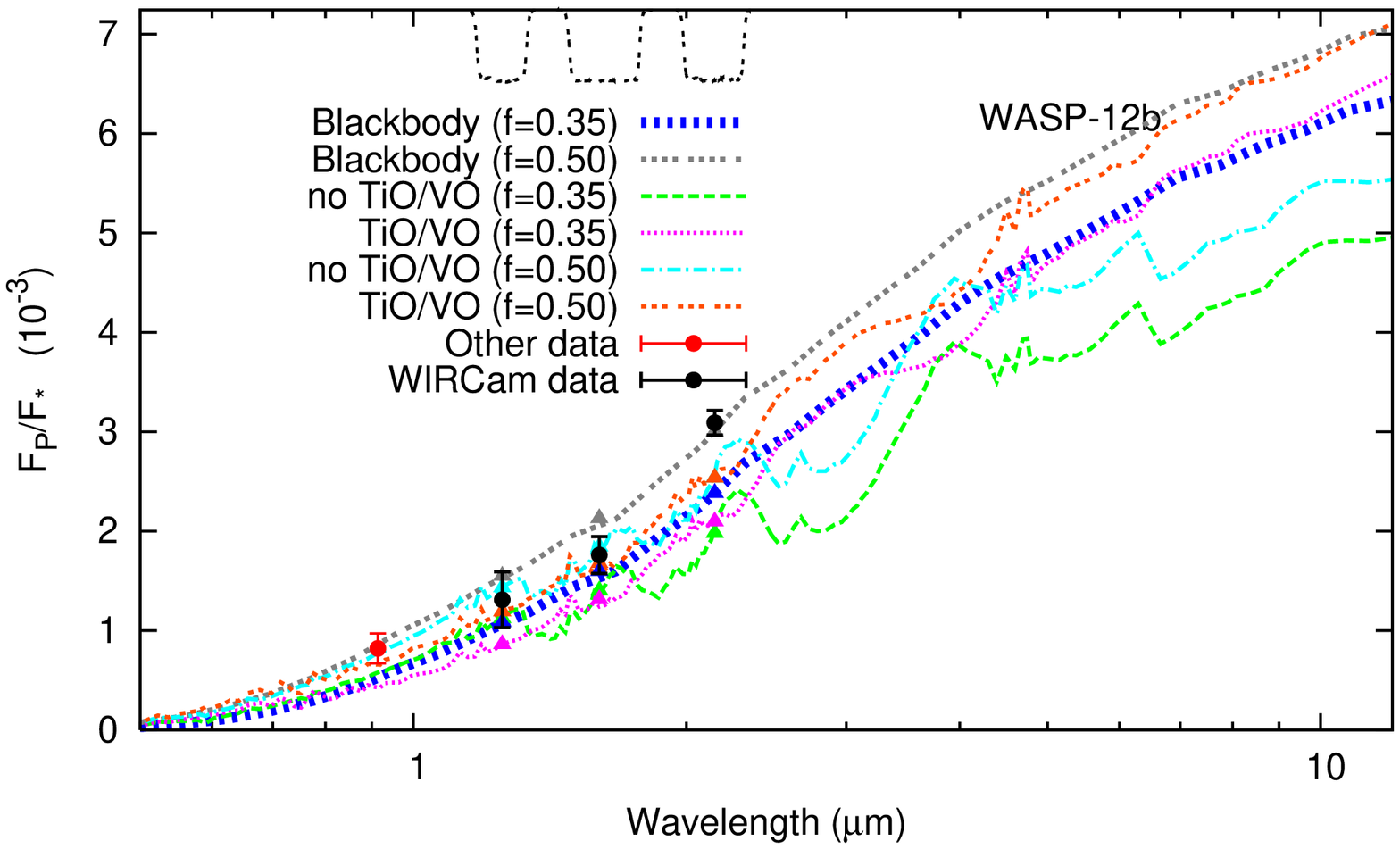}
\includegraphics[scale=0.49,angle=270]{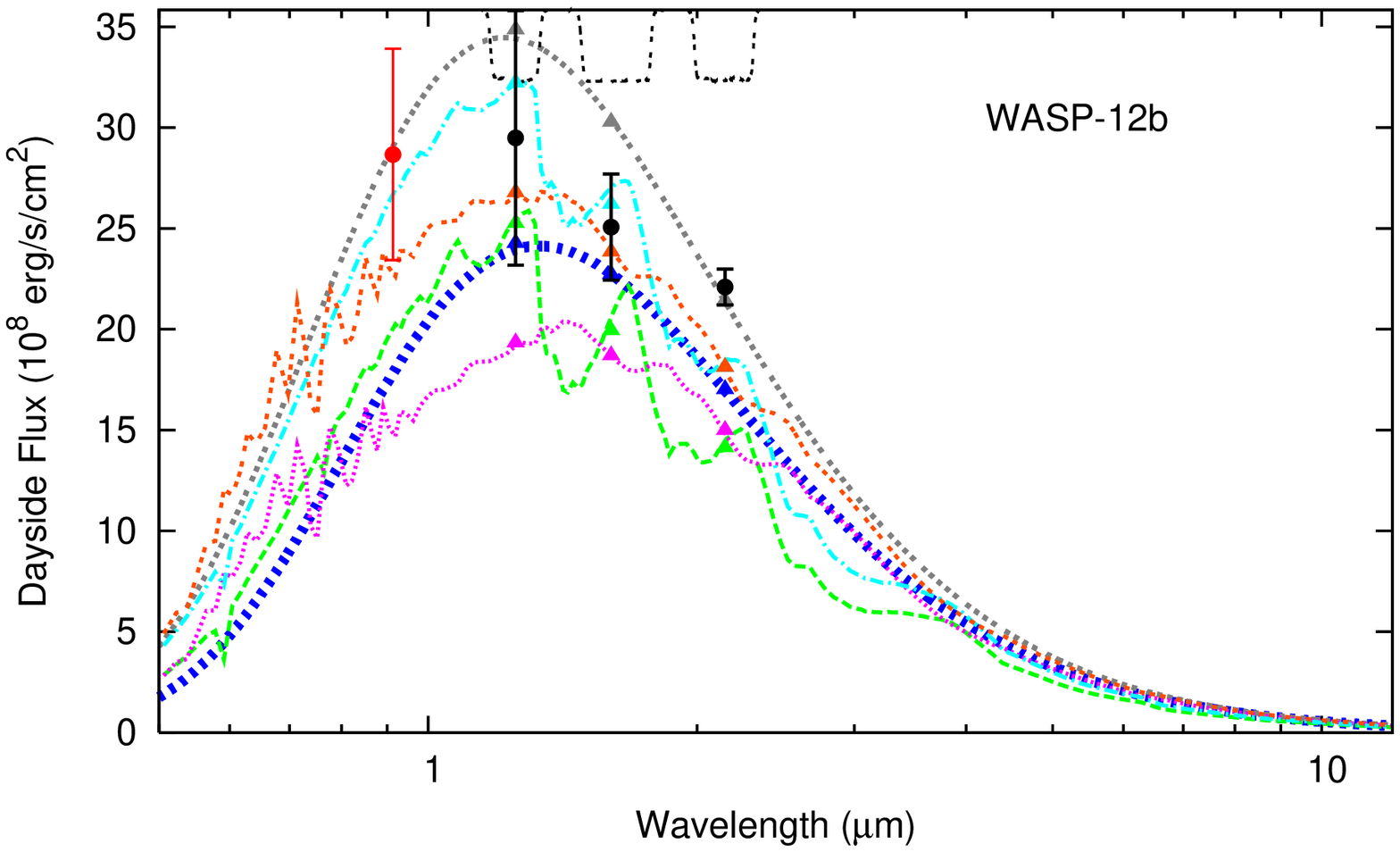}
\caption{	
		Dayside planet-to-star flux ratios (top) and dayside flux at the planet's surface (bottom).
		The Ks-band ($\sim$2.15 $\mu m$), H-band ($\sim$1.60 $\mu m$) and J-band ($\sim$1.25 $\mu m$)
		points are our own, while the z'-band point ($\sim$0.9 $\mu m$) is from \citet{LopezMorales10}.		
		Blackbody curves for modest redistribution ($f$=0.35; $T_{eq}$$\sim$2735 $K$; blue dashed line),
		and dayside only emission ($f$=$\frac{1}{2}$; $T_{eq}$$\sim$2990 $K$; grey dotted line) are 
		also plotted.
		We also plot one-dimensional, radiative transfer spectral models \citep{Fortney06,Fortney08}
		for various reradiation factors and with and without TiO/VO.
		We plot models with modest redistribution ($f$=0.35)
		with and without TiO/VO (magenta-dotted and green-dashed lines, respectively),
		and for dayside only emission ($f$=$\frac{1}{2}$) with and without TiO/VO (orange dotted and cyan dot-dashed lines, respectively).
		The models with TiO/VO display temperature inversions.
		The models on the top panel are divided by a stellar atmosphere model \citep{Hauschildt99} of WASP-12 
		using the parameters from \citep{Hebb09}.
		($M_{*}$=1.35 $M_{\odot}$, $R_{*}$=1.57 $R_{\odot}$, $T_{eff}$=6300 $K$, and log $g$=4.38).
		We plot the WIRCam Ks, H and J-band transmission curves
		inverted at arbitrary scale at the top
		of both panels (dotted black lines). We integrate our models over the WIRCam bandpasses and display the result
		in the appropriately coloured triangles.
	}
\label{FigModel}
\end{figure}

%
% We do not include the \citet{LopezMorales10} eclipse depth in the above analysis due to the aforementioned
% uncertainty with this point. The Spitzer/IRAC eclipse depths \citep{Campo10} are also not included, as of the time of writing only the 
% central eclipse times have been reported; we can expect that the Spitzer/IRAC eclipse depths will significantly increase our understanding
% of this planet's redistribution and reradiation of the extreme incident stellar irradiation. A visual inspection of the \citet{Campo10}
% data indicates that they are consistent with an atmosphere that redistributes little of its heat to the nightside.

	As the atmospheres of hot Jupiters may be highly vertically stratified,
different atmospheric layers may redistribute heat
much more or much less efficiently than other layers. 
The best-fit brightness temperatures and reradiation factors of the individual
atmospheric layers probed by our various wavelengths of observations are:
$T_{B Ks}$=\TBrightWASPTwelveKs$^{+\TBrightPlusWASPTwelveKs}_{-\TBrightMinusWASPTwelveKs}$$K$ and $f_{Ks}$=\fReradiationWASPTwelveKs$^{+\fReradiationPlusWASPTwelveKs}_{-\fReradiationMinusWASPTwelveKs}$ for our Ks-band observations, 
$T_{B H}$ =\TBrightWASPTwelveH$^{+\TBrightPlusWASPTwelveH}_{-\TBrightMinusWASPTwelveH}$$K$ and $f_{H}$=\fReradiationWASPTwelveH$^{+\fReradiationPlusWASPTwelveH}_{-\fReradiationMinusWASPTwelveH}$ for our H-band observations,
$T_{B J}$ =\TBrightWASPTwelveJ$^{+\TBrightPlusWASPTwelveJ}_{-\TBrightMinusWASPTwelveJ}$$K$ and $f_{J}$=\fReradiationWASPTwelveJ$^{+\fReradiationPlusWASPTwelveJ}_{-\fReradiationMinusWASPTwelveJ}$ for our J-band observations.
Our three differents bands
should be probing high pressure regions, deep into the atmosphere of WASP-12b. 
Specifically if the near-infrared opacity is dominated by water vapour opacity the J, H \& K-bands should
be windows in water opacity \citep{Fortney08}, and the Ks, H \& J-bands should be seeing progressively 
deeper into WASP-12b's atmosphere.
Within the errors the brightness temperatures displayed in our three near-infrared bands are similar.
%%ATMOSPHERIC_CHANGE
%Taken at face value the decreased J-band brightness temperature compared to the H and Ks-band
However, the J and H-band brightness temperatures are marginally lower, and taken at face value 
compared to the Ks-band
brightness temperature they suggest a modest temperature inversion at 
very high pressures of $\sim$100 to 500 mbar, deep in the atmosphere of WASP-12b.
%%ATMOSPHERIC_CHANGE
%One explanation for why WASP-12b might display decreased flux in J-band as compared to longer wavelengths,
%is that the atmospheric depths and pressures probed by this shorter wavelength observation 
One explanation for why WASP-12b might display decreased flux at these shorter wavelengths as compared to the Ks-band,
is that the atmospheric depths and pressures probed by these shorter wavelength observations 
may be more homogenized than higher altitude layers. 
The efficiency of redistribution of the incident stellar flux from the dayside to the nightside should
be proportional to the ratio of the reradiative ($\tau_{rad}$) to advective timescales ($\tau_{adv}$).
It is thought that the
reradiative timescale should increase with pressure and 
depth\footnote{The radiative time-scale (how quickly the planet reradiates the incident stellar flux) is thought to be proportional to $\tau_{rad}$ $\sim$ $\frac{c_{P} P}{4 g \sigma T^3}$ \citep{ShowmanGuillot02}, where $c_{P}$ is the specific heat capacity, $\sigma$ is the Stefan-Boltzmann constant, $T$ is the temperature of the atmospheric layer, and $g$ is the gravitational acceleration of the planet.}.
The advective timescale \footnote{It is thought that the advective timescale (how quickly the planet advects the heat to the nightside of the planet; $\tau_{adv}$) can be approximated by the radius of the planet, $R_{P}$, divided by the horizontal windspeed, $U$: $\tau_{adv}$ $\sim$ $R_{P}$/U \citep{ShowmanGuillot02}.}
is also thought to increase in pressure, although it is generally thought that advection should
win out over reradiation as one descends through the atmosphere of a typical hot Jupiter \citep{Seager05,Fortney08}.
Thus, one might expect more efficient redistribution of heat
at the layers probed by our shorter wavelength observations compared to the layers probed by our Ks-band observations.
Other explanations 
for the relatively higher Ks-band emission
than the H and J-band emission are certainly possible, including: 
extra flux from 
a circumstellar disk or material streaming from the planet in the Ks-band, an atmospheric emission feature at Ks-band, 
or absorption features over the H and J-bands.
The eclipse depths from the \citet{Campo10} Spitzer/IRAC measurements will not shed much additional light on this matter,
as, if water opacity dominates, the Spitzer/IRAC bands to do not probe as deeply as the JHK near-infrared bands.
%%ATMOSPHERIC_CHANGE
%at the layers probed by our J-band observations compared to the layers probed by our H or Ks-band observations.
%%ATMOSPHERIC_CHANGE
%We note other explanations for the lower than expected
%J-band emission include an absorption feature at this wavelength, emission features at our other wavelengths, or 
%that we have simply underestimated the systematic errors from observations of this partial eclipse and the true J-band emission
%is greater than we have measured.

	We compare the depths of our near-infrared eclipses to a series of planetary atmosphere models in Figure \ref{FigModel}.
This comparison is made quantitatively as well as qualitatively by integrating the models over the WIRCam J, H \& Ks band-passes
and calculating the $\chi^{2}$ of the thermal emission data compared to the models.
We include the \citet{LopezMorales10} eclipse depth in Figure \ref{FigModel}, but do not include it in our $\chi^{2}$ calculation due to the aforemtioned
uncertainty with the timing and depth of this eclipse. The Spitzer/IRAC eclipse depths \citep{Campo10} are also not included, as of the time of writing only the 
central eclipse times have been reported. 
We first plot two blackbody
models, the first one displaying modestly efficient heat redistribution ($f$=0.35; blue dotted line; $T_{eq}$$\sim$2735 $K$),
while the latter features emission from the dayside only ($f$=$\frac{1}{2}$; grey dotted line; $T_{eq}$$\sim$2990 $K$).
The $f$=$\frac{1}{2}$ blackbody model provides an excellent fit to the longer wavelength
Ks-band emission, and does a reasonable job of fitting our H and J-band emission ($f$=$\frac{1}{2}$: $\chi^{2}$=\BlackbodyTwoChi);
nevertheless it proves a quantitatively better fit than the 
modest redistribution model ($f$=$0.35$: $\chi^{2}$=\BlackbodyOneChi),
which generally underpredicts the observed emission.

	In Figure \ref{FigModel} we also compare our measurements
to a series of one-dimensional, radiative transfer, spectral models 
\citep{Fortney05,Fortney06,Fortney08} with different reradiation factors
that specifically include or exclude gaseous TiO/VO
into the chemical equilibrium and opacity calculations.
In these models when TiO/VO are present in gaseous form in the upper atmosphere
they act as absorbers at high altitudes and lead to 
hot stratospheres and temperature inversions \citep{Hubeny03}.
We present models with modest redistribution ($f$=0.35)
and dayside only emission ($f$=$\frac{1}{2}$) with and without TiO/VO.
The associated $\chi^{2}$ for the $f$=$\frac{1}{2}$
models with and without TiO/VO are $\chi^{2}$=\FortneyFourChi \ and $\chi^{2}$=\FortneyThreeChi, while the $f$=0.35 models
with and without TiO/VO are $\chi^{2}$=\FortneyTwoChi \ and $\chi^{2}$=\FortneyOneChi.
None of these models provide quantitative improvements over the $f$=$\frac{1}{2}$ blackbody model, 
as they do not do as good of job of matching
the longer wavelength Ks-band thermal emission, nor do they feature reduced emission in H and J-band.

 Our near-infrared measurements also allow us to estimate the bolometric dayside luminosity of WASP-12b, $L_{day}$. We use a blackbody model
with a total reradiation factor equal to the best-fit value we calculate from our three near-infrared bands ($f_{tot}$=\fReradiationWASPTwelveALL);
by integrating over this model we can estimate $L_{day}$ as \BolometricFluxDayside$\times$10$^{-3}$$L_{\odot}$.
Another way of parameterizing the efficiency of the day-to-nightside heat redistribution rather than the reradiation
factor is comparing the bolometric dayside luminosity, $L_{day}$, to the nightside luminosity, $L_{night}$. 
By following elementary thermal equilibrium calculations one can deduce that WASP-12b
should display a total bolometric luminosity of 
$L_{tot}$ = \BolometicFluxTotal$\times$10$^{-3}$$L_{\odot}$. This suggests
that \DaysidePercentage\% of the incident stellar irradiation is reradiated by the dayside, leaving a mere \NightsidePercentage\% to be advected to the nightside and reradiated.
However, caution is encouraged with this conclusion as shorter and longer wavelength emission for this planet may deviate 
significantly from that of a blackbody.

\subsection{Future Prospects}

 We lastly note that the combination of 
thermal emission as prominent as that displayed here with near-infrared photometry this precise 
suggests the possibility
that thermal phase curve measurements may be possible from the ground. 
For the shortest period exoplanets
($P$$\sim$1$d$ or less) even in a single night of observing (8-9 hours) one could conceivably view the flux maximum of the phase
curve where hot gas is advected downwind on the planet, the decrement in flux during the secondary eclipse, and then 
view a significant fraction of the near-sinusoidal
decrease as the cool nightside face of the exoplanet rotates into sight.
WASP-12b is an ideal target for such observations with its short 1.09$d$ period, and its bright dayside emission
suggests that thermal phase curve observations for this planet should reveal a large asymmetry over the course of the orbit
as WASP-12b's nightside should be cold.
Thermal phase curve observations from the ground in the near-infrared would require
one to control the background systematic trends that are present in our near-infrared 
photometry even after we correct the flux of our target star with a great many reference stars; the feasibility of this task
is, as of yet, unproven. 
Nevertheless, we will be investigating the possibility of obtaining
such near-infrared phase curve information in this photometry as well as with
future observations of WASP-12b. 
These near-infrared phase curve observations will be accompanied by near-simultaneous, 3.6 and 4.5
$\mu m$ Spitzer/IRAC thermal phase curve observations
of a full orbit of WASP-12b (P.I. P. Machalek) that will allow for an unprecedented understanding of the characteristics of the
day and nightside deep atmosphere of this planet. 

We also plan to reobserve a full, rather than partial, eclipse of WASP-12b in J-band
so as to better define its thermal emission at that wavelength. Lastly we plan to observe the transit of WASP-12b
in the near-infrared Ks and H-bands, combined with our aforementioned
planned reobservations of the eclipse of WASP-12b in these bands. These combined transit and eclipse observations will allow us to confirm
if the Ks-band eclipse is indeed longer in duration than the optical transit, and if so whether this is due to material
tidally stripped from the planet that may or may not form a circumstellar disk in this system.

%We lastly note that the WIRCam Ks-band cuts off at approximately 2.3 $\mu m$. Our Ks-band observations
%are thus nominally sensitive to the 
%2.292 $\mu m$ CO emission predicted to come from the putative
%gas disk around WASP-12b \citep{Li10}. 

% Significant transit timing variation. Eclipse variability due to a storm.

% Comment on why HAT-P-7 is similar. The first detection of the optical albedo of an exoplanet is \citet{Borucki09}.
% we have argued in \citeP{CrollHatP7} that the credit for the first detection of predominantly optical flux
%was from \citet{Borucki09}.

\acknowledgements

The Natural Sciences and Engineering Research Council of Canada supports the research of B.C. and R.J.
The authors thank Daniel Fabrycky, and Darin Ragozzine for helpful discussions on the effects of precession,
and Nicolas Cowan for helpful discussions on the putative disk in this system.
The authors especially appreciate the hard-work and diligence of the CFHT staff
for both scheduling these challenging observations and ensuring these ``Staring Mode'' observations were successful.

%\clearpage
\end{document}